\RequirePackage{fix-cm}
\documentclass[10pt]{svjour3}                     
\usepackage{mathptmx} 

\smartqed  
\usepackage{graphicx}
\usepackage{url}
\usepackage[nomargin,inline,draft]{fixme}
\usepackage{acronym}
\usepackage[section]{placeins} 
\usepackage{dsfont} 
\usepackage{amssymb}
\usepackage{amsmath}
\usepackage[tight,footnotesize]{subfigure}
\usepackage[tight,nice]{units}
\usepackage{float}
\usepackage{mathtools}%
\usepackage{algpseudocode}
\usepackage{setspace}
\usepackage[top=1in,bottom=1in,left=1.25in,right=1.25in]{geometry}

\DeclareMathOperator*{\argmax}{arg\,max}

\DeclareMathAlphabet{\mathpzc}{OT1}{pzc}{m}{it}
\acrodef{RV}{Random Variable}
\acrodef{cdf}{cumulative distribution function}
\journalname{ACM Wireless Networks}
\begin{document}

\title{Location-Quality-aware Policy Optimisation for Relay Selection in Mobile Networks}

\titlerunning{Location-Quality-Aware Relay Selection}        

\author{Jimmy~J.~Nielsen~\and~Rasmus~L.~Olsen~\and~Tatiana~K.~Madsen~\and~Bernard~Uguen~\and~Hans-Peter~Schwefel}


\institute{Jimmy J. Nielsen, Rasmus L. Olsen, and Tatiana K. Madsen \at
              Department of Electronic Systems, Aalborg University, Aalborg, Denmark\\
              \email{\{jjn,~rlo,~tatiana\}@es.aau.dk}           
           \and
           Bernard Uguen \at
              IETR, University of Rennes 1, Rennes, France\\
              \email{bernard.uguen@univ-rennes1.fr}
           \and
           Hans-Peter Schwefel \at
              Department of Electronic Systems, Aalborg University, Aalborg, Denmark, \emph{and}\\
              Forschungszentrum Telekommunikation Wien - FTW, Vienna, Austria\\
              \email{schwefel@ftw.at}
}

\date{Received: date / Accepted: date}

\maketitle


\begin{abstract}
Relaying can improve the coverage and performance of wireless access networks. In presence of a localisation system at the mobile nodes, the use of such location estimates for relay node selection can be advantageous as such information can be collected by access points in linear effort with respect to number of mobile nodes (while the number of links grows quadratically). However, the localisation error and the chosen update rate of location information in conjunction with the mobility model affect the performance of such location-based relay schemes; these parameters also need to be taken into account in the design of optimal policies. This paper develops a Markov model that can capture the joint impact of localisation errors and inaccuracies of location information due to forwarding delays and mobility; the Markov model is used to develop algorithms to determine optimal location-based relay policies that take the aforementioned factors into account. The model is subsequently used to analyse the impact of deployment parameter choices on the performance of location-based relaying in WLAN scenarios with free-space propagation conditions and in an measurement-based indoor office scenario.

\end{abstract}

\section{Introduction}\label{intro}
Two-hop relaying has been shown to improve throughput in WLANs. However, for traditional measurement-based approaches as in References \cite{liu2004raar,zhu2006rdcf,liu2007coopmac}, user movements cause link quality measurements to become outdated and inaccurate, leading to performance degradations. More frequent updates of link measurements can mitigate this degradation, however at the cost of signalling overhead that scales quadratically with the number of nodes.
In Reference \cite{jth2010wcnc}, we considered a location based relay selection scheme that uses estimated node positions provided by a localisation algorithm to determine whether to relay or not and which relay to use. As the signalling overhead scales only linearly with the number of nodes and allows for movement prediction, it is useful in highly mobile scenarios and in scenarios of infrequent uplink transmissions (as would be the case in many M2M communication scenarios with mobile sensors). 

While the simulation models that were used to analyse the location-based relaying scheme in Reference \cite{jth2010wcnc} are sufficient to evaluate the expected performance of a single scenario with a given set of parameters, it is computationally infeasible to use simulations for determining system deployment settings, namely 1) location update rate, 2) optimal relay policies, i.e., in which places to relay or transmit directly, and 3) required accuracy of localisation system.

\begin{figure*}[t]
  \begin{center}
    \includegraphics[width=1\linewidth]{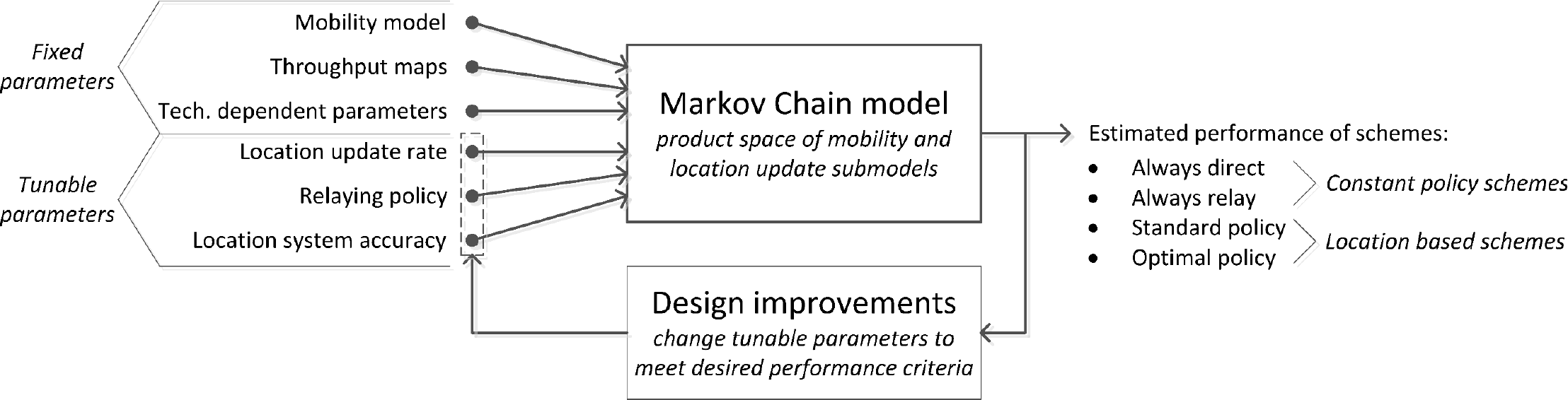}
  \end{center}
  \caption{Intended use of the proposed Markov chain model as a tool for optimizing deployment parameter settings of location based relay systems.}
  \label{fig:mc_model_use}
\end{figure*}

As a consequence, this paper derives a continuous-time Markov chain (MC) model that allows to numerically calculate achievable performance of location-based two-hop relaying while capturing the impact of delays of the information forwarding and of localisation inaccuracies originating from the positioning approach at the mobile nodes. The location information is thereby periodically sent from the mobile nodes to the access point (AP). Such periodic information forwarding represents one of the three basic information access schemes that have been analysed via analytic models in References \cite{bogsted2010probabilistic,olsen2010precise}. The analysis approach of References \cite{bogsted2010probabilistic,olsen2010precise} also inspired the model setup for the relay scenario in this paper.

While there can be substantial further gains of PHY-layer combining of relay transmission and original AP transmission \cite{popovski2007physical}, such approaches require modification of the physical layer of the destination nodes. In order to be able to deploy relaying on top of off-the-shelf physical layers, this paper considers store-and-forward relaying without additional Layer-1 symbol combining. Notice however that the contribution of this paper does not lie in the used relaying scheme, but rather in the MC modelling framework and optimization algorithms. The considered store-and-forward relaying scheme is merely a use case example. In principle, other more advanced relaying schemes can be used with the proposed model, as long as appropriate location dependent utility functions can be defined.

The diagram in Fig. \ref{fig:mc_model_use} shows an overview of how the proposed MC model can be used to determine deployment parameters: While the mobility model of a node, the technology dependent channel characterization, and parameters of the wireless technology (upper left) are considered given by the scenario, the location update rate, the actual relaying policy and the choice of the location system (lower left) can be influenced by the deployment configuration. The Markov model and the algorithms in this paper provide a method for optimizing this configuration.

Location-based relay selection was considered already in References \cite{zorzi2004geographic,zhao2005harbinger,yang2008novel,yang2009extending} and for cooperative MIMO scenarios in Reference \cite{nasser2009positioning}, however in all cases under the assumption of perfect non-delayed location knowledge.
The impact of feedback delay on relaying has been studied in, e.g., References \cite{suraweera2010performance,seyfi2011performance}, but in both cases for systems that are not utilizing location information.
Markov based models have been used for mobility modelling, location tracking and trajectory prediction as in References \cite{liu1998mobility,liao2007learning,jnt2013ew}. The mobility models from these papers can be used to describe the movements of the mobile node(s).
To the best of our knowledge only our own previous work in References \cite{jrth2012wcnc,jrtbh2012camad,jrth2013vtc} uses models of location error, information collection and mobility models to optimize relay decisions.

The Markov model for the analysis of location-based relaying performance was first introduced in Reference \cite{jrth2012wcnc}; subsequently, it was applied to a case study using indoor measurements in Reference \cite{jrtbh2012camad} and it was extended with an efficient policy optimization algorithm and applied to a mobile destination scenario in Reference \cite{jrth2013vtc}. The present paper generalizes and extends those previous modelling approaches and provides the following additional results:
i) The location estimation error is introduced in the model-based analysis and in the evaluation results. The new model therefore allows to consider the joint impact of errors of the localisation system and inaccuracies caused by mobility in conjunction with information access delays.
ii) The algorithm for efficient policy optimization targeting maximization of average throughput presented in Reference \cite{jrth2013vtc} is generalized so that it can be applied to scenarios with mobile destinations as well as mobile relays. Furthermore, the generalized algorithm addresses also cases with multiple relays. 
iii) The measurement-based case study originally introduced and studied with heuristic policies in Reference \cite{jrtbh2012camad} is analysed in scenarios that consider now also location estimation error (see Item (i)) and a rigorous policy optimization is performed and analysed. 
iv) The extended and generalized model of this paper allows to quantitatively analyse the required accuracy level of localisation solutions such that they are beneficial for location-based relaying scenarios. 

The paper is structured as follows: Section \ref{sec:system_description} introduces the general relay system and the collection of location information; the corresponding Markov models are described in Section \ref{sec:mc_model}. Section \ref{sec:performance_metrics} introduces the method for calculating the considered performance metrics. The proposed policy optimization that is generalized to work with one or more relays is presented in Section \ref{sec:opt_policy_calc}. Hereafter two case studies and corresponding performance results are presented: first an outdoor open field scenario in Section \ref{sec:case_study_1} and subsequently a realistic indoor measurement-based case study in Section \ref{sec:measurement_case}. Finally, Section \ref{sec:conclusion} presents conclusions and outlook.




\section{System Description}\label{sec:system_description}
\begin{figure}[t]
\centering
    \subfigure[Mobile relays]{
        \includegraphics[height=4cm]{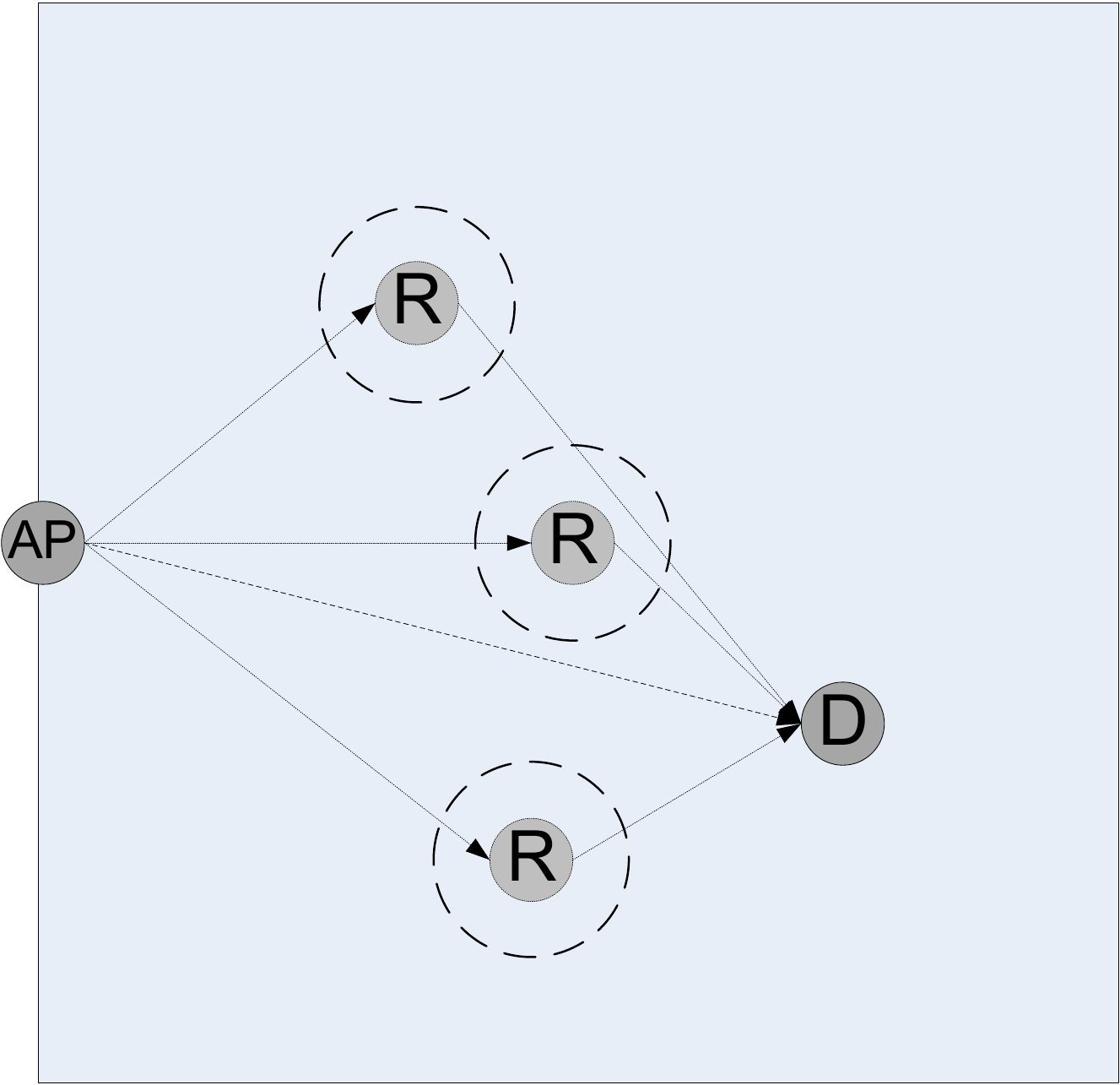}
    \label{fig:3-node-system-mob-relay}
    }
    \subfigure[Mobile destination]{
        \includegraphics[height=4cm]{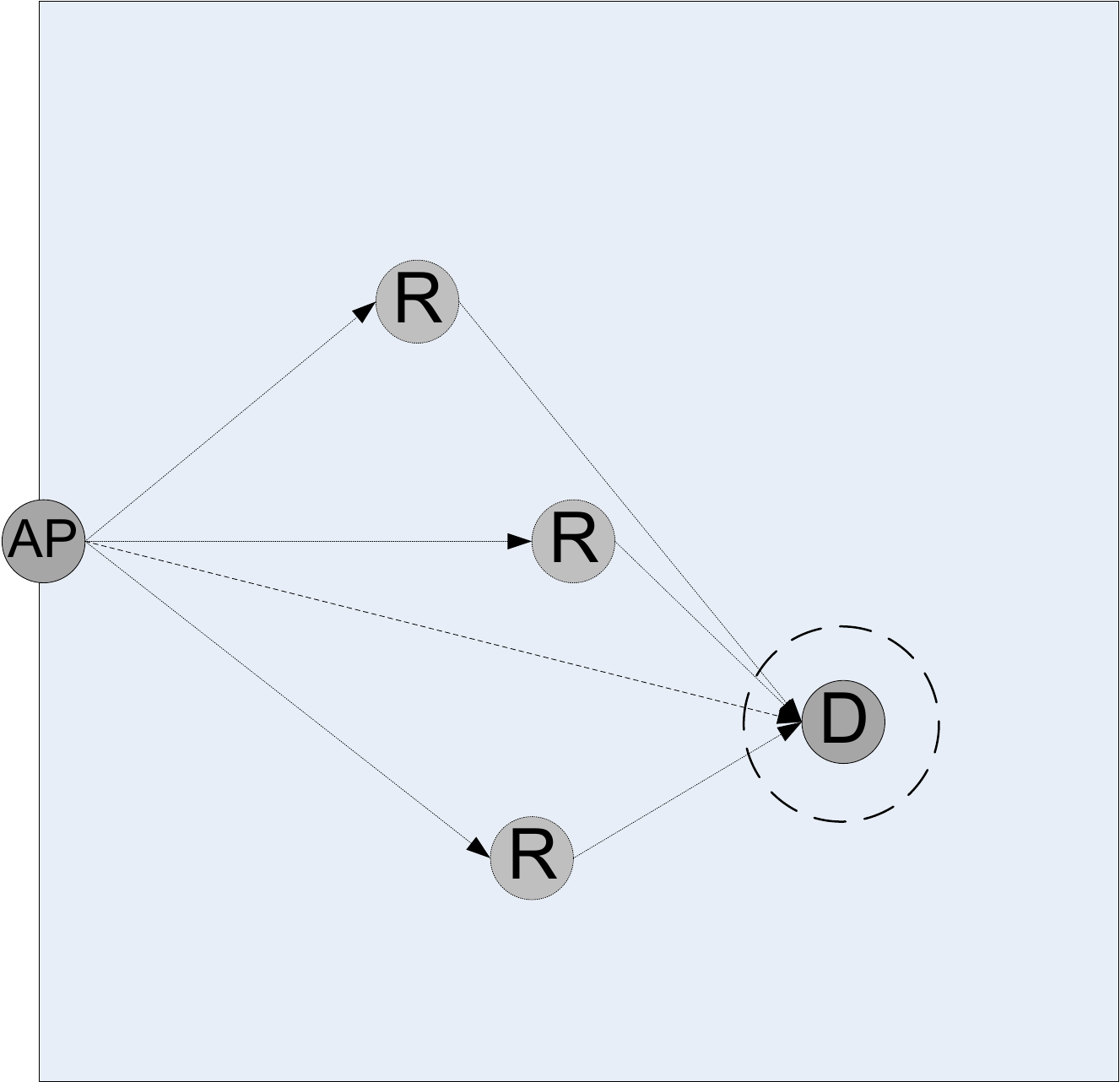}
    \label{fig:3-node-system-mob-dest}
    }
    \caption{System with static AP, relays (R), and destination (D). Position uncertainty for mobile nodes is shown by the dashed circle.}
\label{fig:3-node-system}

\end{figure}

We consider downstream communication between a stat\-ic access point (AP) and a destination node (D); there is a set of $K$ nodes in the geographic region that can potentially act as relay nodes, see  Fig.
\ref{fig:3-node-system}.  Mobility is present either for the destination node or for the relay nodes. Even though multiple mobile nodes can be present, we for simplicity in the following refer to the Mobile Node (MN), which has a position $\mathpzc{x}(t)$ that
changes over time. Via a positioning system (e.g. GPS based), the MN can obtain an estimation of its own coordinate, labelled $\tilde{\mathpzc{x}}(t)$. 
In a location-based relaying approach, the MN will send this coordinate estimate to the AP, so that the AP has a view  $\hat{\mathpzc{x}}(t)$ of the MN's coordinate, which in the general case is a sampled and delayed version of $\tilde{\mathpzc{x}}(t)$. The AP then will take a decision
based on this inaccurate and delayed knowledge of the position of the mobile node: which of the candidate relay nodes to use or if it is better to make a direct transmission to the destination. The mapping of
the position of the MN to the relay decision is called a {\it relay policy}, here,  for the example of a mobile destination node, represented by
$\pi(\hat{\mathpzc{x}}) \in \{R_0=D, R_1,..., R_K\}$, where $R_i$ stands for
'relaying via node $i$' while $R_0=D$ stands for a direct transmission. In this
paper, we assume that this relay decision is taken by the AP just before
the next individual downstream data fragment transmission. The policy optimization algorithms presented later in this paper are used to determine the optimal relay policy for the AP, given the present scenario conditions.

\begin{figure}[t]
\centering
\includegraphics[width=0.325\textwidth]{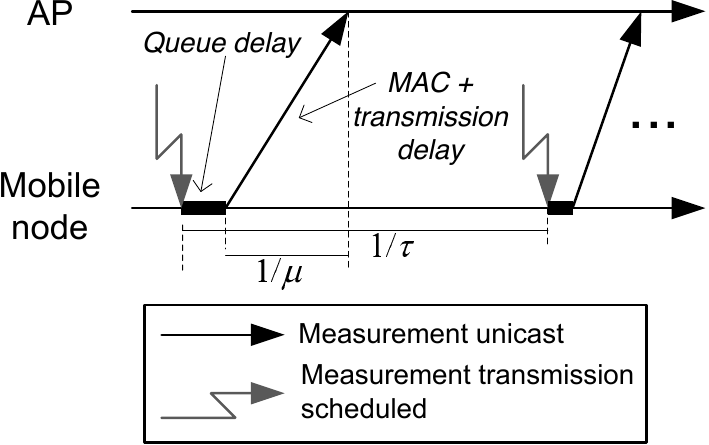}
\caption{Location measurements are transmitted periodically with rate $\tau$ and are subject to queuing delay as well as MAC (Medium Access Control) and transmission delay, represented by a distribution with mean $1/\mu$.}
\label{fig:msc_collection_loc}
\end{figure}

The positions of the static AP and of the static nodes are assumed to be known at the AP,
hence only the MN will periodically (with rate $\tau$) send position updates to the AP.
It is assumed that such location information is available at the MN through, e.g., a GPS system, however the coordinate provided by the chosen positioning solution in the MN can show a stochastically varying error. Furthermore, the forwarding delays of positioning information in scenarios with mobile nodes lead to additional inaccuracies of the position estimate. Investigating the impact of location estimation errors and location information collection delay on the system performance is key to this paper.

While the model in this paper can be applied to different wireless technologies, we use the terminology and procedures subsequently from WLAN 802.11 type transmissions.

The location measurements are transmitted from the MN to the
AP as sketched in Fig. \ref{fig:msc_collection_loc}: 1) the MN 
obtains a, potentially erroneous, estimate of its current location, 2) wraps it into
a WLAN packet and passes it to its WLAN interface. At the WLAN
interfaces, there could be a queuing delay until the location
message reaches the first position in the (finite) interface
queue, 3) followed by a subsequent MAC and transmission delay. The
sum of MAC and transmission delays are assumed to show a
distribution with mean $1/\mu$. The location update message can also be lost, either due to a full WLAN interface queue, or due to an unsuccessful WLAN transmission. The latter happens with probability $p_\text{loss}$. Since the location update messages are very small in size, they can justifiably be transmitted directly from the mobile node to the AP, since a low bit-rate and hence robust modulation scheme can be used. Relaying of transmissions is thus only considered for data transmissions with much larger payload, where it is desirable to achieve as high as possible bit-rate.

The AP's estimate of the MN's position is based on the last received
location measurement. Typically the location estimation error of such measurements from for example satellite or local radio-based location systems is assumed to follow a 2D Gaussian probability distribution \cite{gustafsson2005mpu,sayed2005nbw,mao2009localization}. Outdoors, a zero mean distribution is often assumed, however indoors or in situations with shadowing effects caused by buildings, walls, furniture or other obstructions, the distribution may be offset in some direction or even be non-Gaussian. As the relay decision depends on the location measurement, its location error is therefore an important influencing factor in the system. It is important to emphasize that we do not compare performance of different location systems, but rather we focus on location errors as input to the system.
Even for an ideal positioning system at the MN, since the MN is mobile, its true position may
differ from the AP's estimate, depending on the stochastic mobility model of the MN. 

Depending on the AP's belief on the MN's location it
will choose to either make a relayed data transmission through one of the $K$ candidate relay nodes, $R_i$, or a direct data transmission (D). It will use the pre-computed relaying policy to determine which transmit mode is expected to yield the highest throughput for the MN's believed location. The resulting achieved throughput will depend on
this choice since the bit rate can be adapted to the quality of each used link in either the direct or two-hop transmission. For simplicity and in order to allow relaying approaches to be implemented on top of existing Layer 1 and 2 implementations, in this study, we do not consider joint decoding of the first and second transmissions in case of relaying; however,  the later introduced throughput models can be extended to take into account such advanced PHY-layer combining approaches. 

For optimal performance, it is desirable to make the choice that maximizes the overall achieved throughput. In the considered scenario, this optimal choice depends on the MN's mobility model, the accuracy of the localisation system, on the distance-dependent propagation characteristics, and on the strategy (period of the location updates) and forwarding delays (queuing, MAC and transmission) of these location updates. The next section develops a Markov model that subsequently will be used for evaluating relay policies and in the calculation of relay policies that maximize throughput.



\section{Markov Model for Single Candidate Relay Node}\label{sec:mc_model}
In order to evaluate relaying performance and to later on determine optimized location-dependent relay policies, this section provides a Markov model that captures the localisation and location update procedures between mobile node and central access point. Main target of the model is to include the localisation errors originating at the positioning system and to capture the impact of information forwarding delays on the location knowledge at the access point. The Markov model in this section considers one destination node and a single relaying node, where one of the two is mobile (called Mobile Node, MN). Whether it is the destination node or the relay node that is mobile, is in fact irrelevant for the model in this section. 

In short, the model in this section takes a mobility model, an information forwarding model from the MN to the AP, a location error model and a location-dependent relay policy as input; the Markov model allows to compute state probabilities of the true coordinates conditioned on a relay or direct transmission decision at the AP; the AP looks up the decision from a given policy based on the resulting delayed and inaccurate location knowledge. The subsequent section then shows how, together with a geographic throughput model as input, resulting performance metrics for this input policy can be calculated in the single relay scenario. Based on this performance model, Section \ref{sec:opt_policy_calc} then develops a computationally efficient (polynomial time with respect to size of geographic region) approach to calculate optimized policies that even can be applied to scenarios of $K$ relay nodes. The key idea in Section \ref{sec:opt_policy_calc} is to use 'singular relay policies' as input to the model of this section in order to compute the conditional probability of the mobile node's true location  $\mathpzc{x}$ conditioned that the assumed location of the MN at the AP is $\hat{\mathpzc{x}}$. These conditional probabilities can then be utilized to determine optimal choices in the presence of $K$ relay candidates; only in that last step, the differences between the scenarios of the relays being mobile or the destination being mobile will become relevant.  

As a first step, this section develops a Markov model that captures the node mobility, the localisation error, and the information forwarding of the positioning information to the access points. It consists of two parts: 1) a continuous-time Markov model for the spatial mobility of the MN (the 'true' coordinates); 2) a model of location update procedures and of the resulting AP view. As these two parts are not completely independent, the transition rates in this product space require subsequent modifications, which then will depend on the positioning error model as well as on the relaying policy, as described in this section.

\subsection{Markov Mobility Model}\label{sec:markov_mobility}
\begin{figure*}[t]
\centering
    \subfigure[Mobility model]{
        \includegraphics[width=0.4\linewidth]{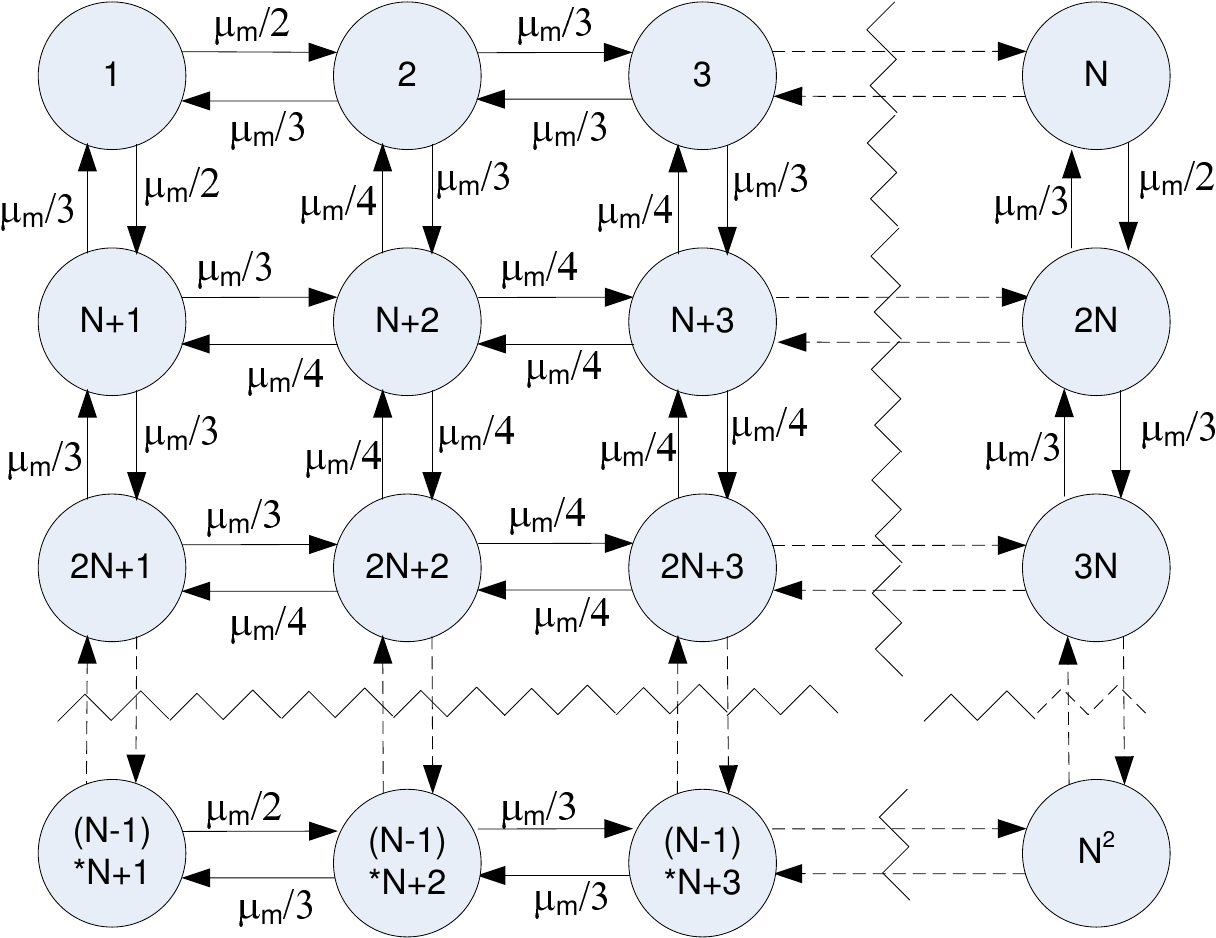}
    \label{fig:mobility_model_nowall}
    }\quad\quad
    \subfigure[Mobility model with wall]{
        \includegraphics[width=0.4\linewidth]{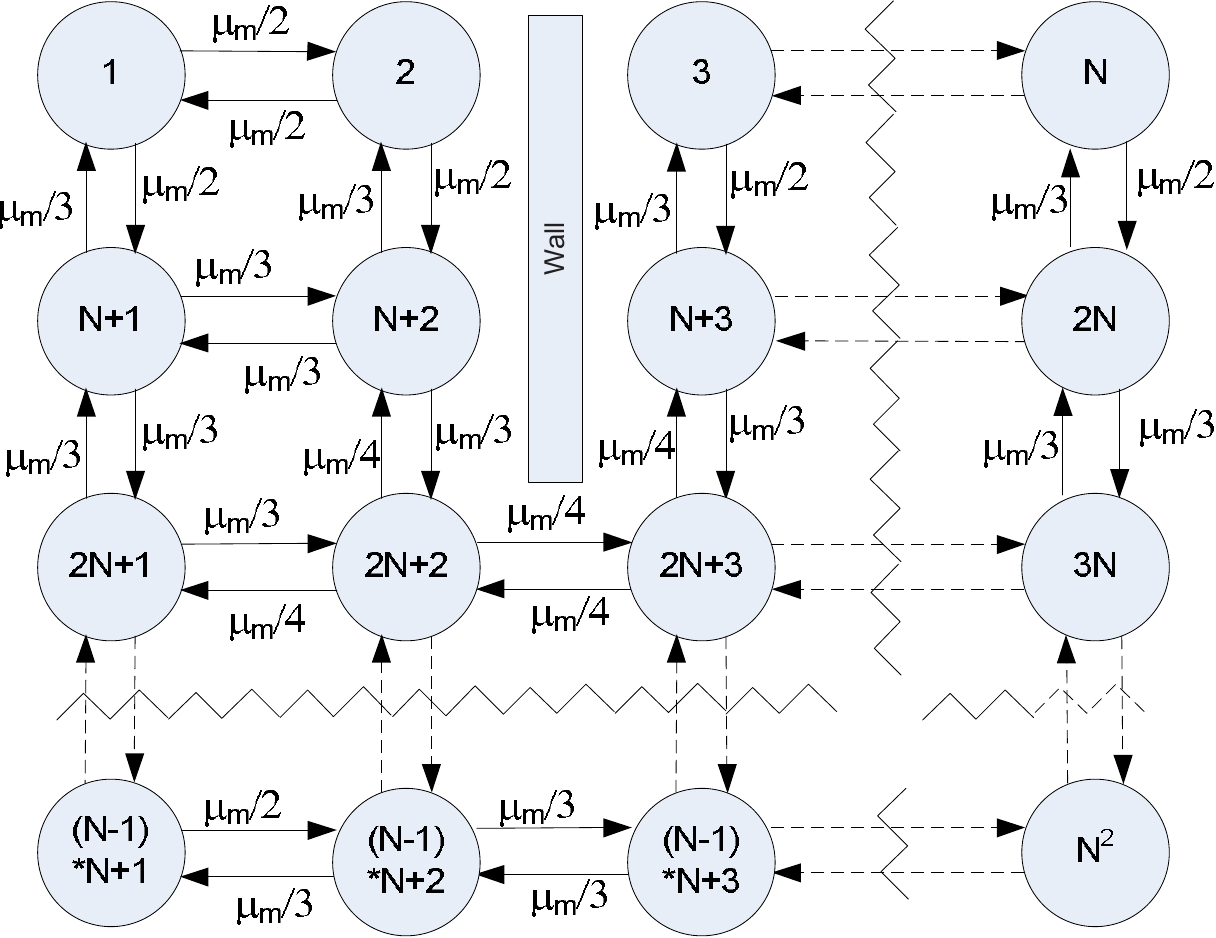}
    \label{fig:mobility_model_wall}
    }
    \caption{(a) Example Markov mobility model that is used in numerical results. In total $N$ discrete grid-points represent the geographic space. Although more general models are possible, this example uses an equidistant grid and one state per grid-point. Also, the state-leaving rate is always $\mu_\text{m}$, which facilitates the computation of an average speed parameter.
(b) A similar mobility model, but with a wall.}
\label{fig:mobility_model}
\end{figure*}

First element of the relaying Markov model is a continuous-time Markov model that describes 
the MN's stochastic mobility. The geographic 2-dimensional space is discretised, 
for instance via an equidistant grid. The states then represent the current true position of the 
MN within the grid. Transition rates between the states characterize the mobility. 
Fig. \ref{fig:mobility_model_nowall} shows a base model without obstacles: 
Transitions are only allowed to the neighbouring grid states and all states have the same overall state 
leaving rate $\mu_\text{m}$. As a consequence, the average movement speed of the candidate relay 
can be readily obtained as $\bar{v}=d/\mu_m$, where $d$ is the distance between neighbouring grid-points.

While the considered mobility model is a special case of user mobility, its advantage is that it only has few parameters, which facilitates conclusions from parametric studies that are performed later in the paper.
Note that more general mobility models can be utilized (see also Reference \cite{olsen2010precise}), in particular along the following lines: 1) states can be associated with any discretisation of the geographic space (so equidistant placement is not needed); 2) transition rates and transition structure can be arbitrary (though for physical movement resemblance, typically transitions would only target geographically neighbouring states); 3) multiple Markov states can be utilized for each discrete coordinate in order to keep memory of directional information (as in Reference \cite{jnt2013ew}) or to mimic non-exponential state-holding times via Phase-type distributions. Figure \ref{fig:mobility_model_wall} shows an example mobility model for indoor scenarios, in which a wall blocks certain movements.

The generator matrix of the Markov mobility model is in the subsequent sections denoted as $Q_\text{mob}$ and a mapping function of the state number $m$ to geographic coordinate is used: $c: \{1,2,...,N^2\} \rightarrow \mathbb{R}^2$. Inversely,  $(x_i)_{i=1,...,N^2}$ denotes the enumerations of the $N^2$ coordinates of the Markov mobility model.
The steady state probability distribution of the mobility model is denoted by $p^\text{mob}$. 
In addition to the regular grid model, a more complicated mobility model with obstacles will be used in the measurement-based case study; this model will be explained later in Section \ref{sec:measurement_case}.

\subsection{Relay Policy Representation for Single Relay Case}
Before we model the information forwarding in the next subsection, we introduce here the notion of a relay policy for the single-relay-candidate case, which will be input to the information forwarding model. Later, more general policies and more general cases will be discussed. The simple case of one candidate relay node enables a compact model formulation, as it allows to reduce the state-space of the models, since only the decision (D or R transmission) and not the full geographic coordinate space needs to be encoded in the relevant Markov model states. This single-relay case will then later be the instrument to address more complex scenarios in a computationally efficient manner. 

This simple single-relay-candidate policy is a function that maps geographic coordinates of the MN to the binary relaying decision, which is either a direct ('D') or relayed ('R') transmission. Hence, the single-relay policy is a function $\pi: \hat{\mathpzc{x}} \rightarrow \{R,D\}$. Note that the policy is implemented at the AP, hence it uses the estimated MN position as input, i.e. both positioning errors of the location system as well as forwarding delays will impact the reliability of this information. The modelling of this position estimate at the AP is described in the next section.

\subsection{AP View and Information Forwarding}
\begin{figure}[t]
    \centering
        \includegraphics[height=11.5cm]{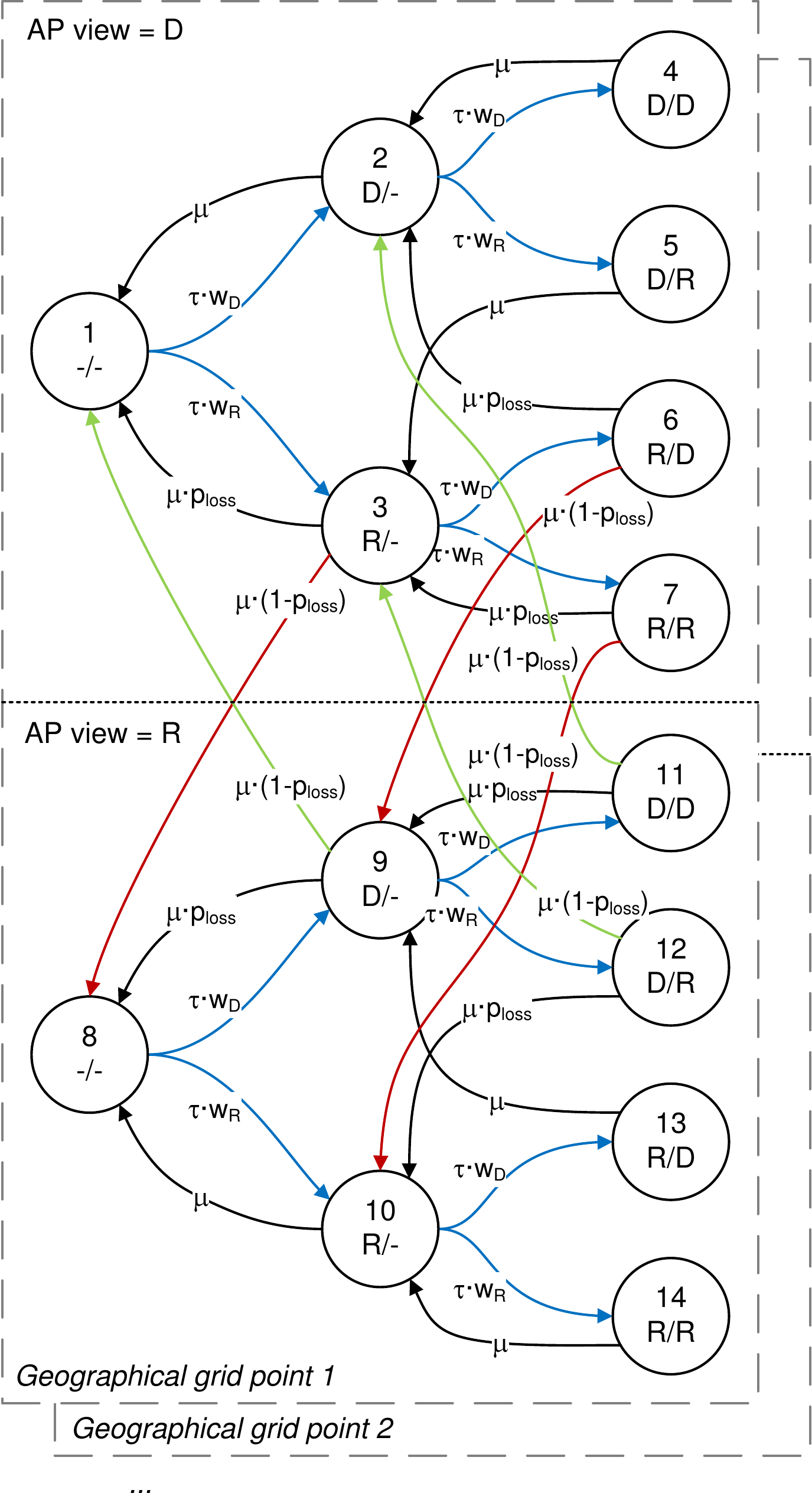}
    \caption{State overview for Markov model of information forwarding, here depicted for queue length $N_\text{q}=2$, which is also used for the later numerical studies.}
    \label{fig:state-overview}
\end{figure}

In order to model the AP view on the location information resulting from the location update process, the state-space at each coordinate (state of the mobility model) is extended as illustrated in Fig. \ref{fig:state-overview}. The model assumes a two-element WLAN interface FIFO queue at the MN. First, memory of the AP on the last received coordinate needs to be introduced. Instead of keeping track of coordinates, it is sufficient to keep track of the relay decision $\pi(\mathpzc{x})$ that corresponds to that last communicated coordinate. Hence, the AP can be in state 'D' (last received location update was a position that is mapped to a direct transmission in the relaying policy; upper half of Fig. \ref{fig:state-overview}) or state 'R' (analogous for a relay transmission; lower half of Fig. \ref{fig:state-overview}). When the MN triggers a location update (with rate $\tau$),  the state-space does not need to encode the actual full coordinate contained in this location update, rather it is enough to encode the resulting decision $\pi(\mathpzc{x})$.
If the current state of the mobility model (true coordinate) is associated with $\pi(\mathpzc{x})=$'D', then the update in progress is memorized as such (State 2 and 9 in Fig. \ref{fig:state-overview}; otherwise States 3 and 10). As soon as the update is received at the AP, the AP's view is changed accordingly, leading for instance to the transition from State 3 (AP view is 'D', while coordinate update leading to 'R' is in progress) to State 8 (AP view is 'R', no update in progress). 

When queuing of state-update messages at the WLAN interface queue may occur, additional states are needed: In the shown example, the max queue-size is set to 2 (one update in progress of being transmitted, while one can be in the buffer). States 4-7 and 11-14 thereby correspond to the situation that one update is being transmitted, while a second one  is stored in the queue. The state label describes the content of the location updates, where the first element refers to the location update in progress. 
Increasing the maximum queue size is simply done by adding two additional states for each of the right most states, i.e. in Fig. \ref{fig:state-overview} that would be state 4-7 and 11-14 and connect them in a similar way. The state space for each AP view would contain $2^{N_\text{q}+1}-1$ elements ($N_\text{q}$: queue-size), and thus, the whole state space size is $2(2^{N_\text{q}+1}-1)$ for a single geographic grid point. For the complete grid model with $N^2$ geographic grid points, the complete state space model contains $(2(2^{N_\text{q}+1}-1))N^2$ states.

Measurements are transmitted according to the measurement transmission delay rate $\mu$. Finally, potential losses of update messages during the wireless transmissions are also modelled; they occur with probability $p_\text{loss}$. Note that losses of updates due to full interface queues are modelled implicitly, as the updates with rate $\tau$ are not occurring any more in the rightmost states. Notice that the Markov model allows the measurement loss probability and measurement transmission rate to be location-dependent. However, for simplicity and because the small packet size would justify using a more robust modulation scheme than for data transmissions, these parameters are in this paper considered to be independent of location. 

In order to express the relay policy's dependence on the location of the MN, i.e., whether a 'D' or 'R' measurement is sent from the MN, the $\tau$ transitions in Fig. \ref{fig:state-overview} are weighted by the binary variables $w_\text{D}(\mathpzc{x})$ and $w_\text{R}(\mathpzc{x})$, where $w_\text{R}(\mathpzc{x})=1-w_\text{D}(\mathpzc{x})$. $w_\text{D}(\mathpzc{x})=1$ if the MN's current true position maps to a D in the relay policy, else $w_\text{D}(\mathpzc{x})=0$.
When introducing location error in the next subsection, the variables $w_\text{D}(\mathpzc{x})$ and $w_\text{R}(\mathpzc{x})$ will represent the probability that the true location is mapped into a Direct or Relay decision, hence these factors will become real values between 0 and 1, while they at single coordinate always add up to 1.



The depicted 14 states in the example shown in Fig. \ref{fig:state-overview} exist for each grid point in the mobility model. The transitions between grid points in the mobility model, shown for the considered example in Fig. \ref{fig:mobility_model}, are independent of the states of the AP view and information forwarding; hence they do not lead to changes of the corresponding state in Fig. \ref{fig:state-overview}.
%

\subsection{Modelling of Location Errors}
The location error is modelled as a probability distribution that describes the outcome of the location system, $\tilde{\mathpzc{x}}$, given a certain true position, $\mathpzc{x}$ of the node. Notice that this location error only reflects inaccuracy of location estimation, not the inaccuracy caused by collection delays.
Using the discretised geographic space which is the basis for the Markov mobility model and given the enumeration $(x_i)$ of all possible geographic coordinates, the location error is given by the error matrix $E$ with elements: 
\begin{align}
E_{i,j} = Pr(\mbox{positioning system provides coord. } x_j | \mbox{true coord. is } x_i)
\end{align}
In case of an ideal positioning system, $E$ is the identity matrix.

In relation to the full Markov chain model, the state-space of the Markov chain remains unchanged as it represents the true location of the MN. The error matrix will however influence the mapping to relay choices as follows: For a given true coordinate $i$, the factor $w_R$ in the example of a single candidate relay of Fig. \ref{fig:state-overview} is calculated as the sum of the probabilities in the location error matrix, for which the policy  $\pi$ dictates a relayed (R) transmission:
\begin{align} 
w_R(x_i) = \sum_{j \text{ with } \pi(x_j)=\text{'R'}} E_{i,j}.
\end{align}
Equivalently for $w_D$, the sum is taken over all probabilities $E_{i,j}$ where $x_j$ is a coordinate mapped to a Direct transmission.

\section{Performance Metrics}\label{sec:performance_metrics}
The complete Markov chain model, which combines the Markov chain mobility model in Fig.~\ref{fig:mobility_model} and the information forwarding Markov chain in Fig.~\ref{fig:state-overview}, is created by the following two-step procedure.
\begin{enumerate}
  \item Create a full generator matrix template $Q^*$ as the Kronecker product of the mobility model generator matrix $Q_\text{mob}$ and information forwarding generator matrix $Q_\text{info}$:
  \begin{equation}
  Q^* = Q_\text{mob} \otimes Q_\text{info}
  \end{equation}
  \item Adjust location dependent transition rates, by iterating over states and inserting the corresponding values. Specifically, in the presented case studies, the location dependent rates that we update are the $\tau \cdot w_R$ and $\tau \cdot w_D$ transition rates. Hereby, we get the final generator matrix $Q$.
\end{enumerate}

Having set up the Markov chain model with generator matrix $Q$ for a single-relay policy as input, the steady-state probabilities $p$ of this Markov chain can be obtained from the linear equation system:
\begin{align}
p Q = 0, \quad \sum_i p_i = 1.
\end{align}
Using standard numerical methods for solving linear equations, this step has time complexity $O((N^2\cdot L)^3)$ where $N^2$ is the number of geographic states utilized in the mobility model, and $L$ represents the number of states utilized for the information forwarding process ($L=14$ in the above illustrated example of a interface queue at the mobile node with two places). On a standard PC, for the case studies presented in this paper where the number of grid points is in the range of 100-150, the calculation of steady state probabilities takes up to a few seconds. In relation to this we note that it only takes a few minutes to determine the optimal relay policy for a specific scenario using the algorithms presented later in this paper. From this we conclude that the proposed modelling approach, without specifically optimized code implementation, is feasible in situations where the area dimensions combined with the desired grid resolution results in up to some hundreds of grid points. As comparison, note that the office building case study in Section~\ref{sec:measurement_case} has 147 grid points.

These steady-state probabilities can now be used to numerically calculate different performance metrics for this single-relay case with given input policy. Note that so far we have not distinguished yet, whether the mobile node is the candidate relay node or the destination node. So both cases from the introduction are still covered by the model.
For increased readability, we index the $p$ vector corresponding to the product space representation, i.e.  $p_{m,s}$ refers to the $s$th state of the information update model model in the $m$th grid point of the mobility model. 

\subsection{Throughput-Based Performance Metrics}

The steady-state probabilities of the above Markov chain solution allow to compute metrics such as the probability of occurrence of a wrong relay decision. In order to analyse the consequence of such wrong decisions, we need to enrich the model's input with a propagation model, more specifically here, a throughput model.

\paragraph*{Geographic Throughput Model}\mbox{}\\
In order to calculate expected throughput of the relaying system, we here assume that this expected throughput between AP and destination node is only influenced by variability due to changing coordinates of the participating nodes; statistical variations due to changing propagation environments are not considered. Hence, the model uses a mapping of node coordinate pairs to link throughput, respectively for the case of relaying, a mapping of node-coordinate triplets to throughput of the 2-hop link. As we here assume that AP and either destination or relay nodes are static and known, we only require the MN's position as input; motivated by the discretisation of the geographic space, we utilize the state-number rather than the geographic position as input to these throughput functions. Hence, we use two functions, one  $T_\text{D}: (m) \rightarrow R_0^{+}$ for the direct transmission, and another $T_\text{R}: (m) \rightarrow R_0^{+}$ for the relayed transmission.  The specific choice of these throughput functions will now depend on the scenario; for instance, when the mobile node is the relay candidate and the destination node is fixed, $T_D$ will be a constant.

The specific throughput models used for each of the two considered case studies in the mobile relay scenario are described separately in Section \ref{sec:case_study_1} and Section \ref{sec:measurement_case}, respectively.

\subsubsection{Mean Throughput of Relay Policies}
The mean throughput that is achieved via a certain relaying policy can be computed by the weighted sums of the achievable throughput in the correspondingly chosen relay mode; the weights are thereby the steady-state Markov chain probabilities. An ideal relay selection would pick the relay node or the direct transmission that maximizes throughput for the current true position. In practice, this would require an ideal positioning system at the MN, a zero transmission delay and an infinite location update rate. As a comparison case, it can be computed from the Markov model  as follows: 
\begin{align}
    S_\text{ideal}(\pi_\text{opt,ideal})  &= \sum\limits_{m=1}^{N^2} p_m^\text{mob} \cdot
        \max(T_\text{D}(m),T_\text{R}(m))
\end{align}
The achievable throughput from policy $\pi$ under the mobility and information forwarding conditions of the model as described in the previous section is then computed as:
\begin{align}
    S_\text{loc}(\pi) = \sum\limits_{m=1}^{N^2} \Big( &\sum\limits_{s \, \in \, \text{'D' states}}  p_{m,s} \cdot T_\text{D}(m) + \sum\limits_{s \, \in \, \text{'R' states}}  p_{m,s} \cdot T_\text{R}(m) \Big)
  \end{align}

which reduces for the special cases of policies that always transmit directly (dir) or always transmit via relaying (rel) to:
\begin{align}
    S_\text{dir} &= \sum\limits_{m=1}^{N^2} p_m^\text{mob} \cdot T_\text{D}(m)\\
    S_\text{rel} &= \sum\limits_{m=1}^{N^2} p_m^\text{mob} \cdot T_\text{R}(m).
\end{align}

The state set \textit{'D' states} refers to all Markov states at a certain coordinate, in which the AP view is 'D' (in the example of Fig. \ref{fig:state-overview}, this is the upper half of states with IDs $1,\ldots,7$). Analogously for the \textit{'R' states}.
$T_\text{D}(m)$ and $T_\text{R}(m)$ are the expected
throughput with the Mobile Node being at the $m$th grid point for direct
and relayed transmissions, respectively. For constructing the
throughput functions $T_\text{D}(m)$ and $T_\text{R}(m)$, the
throughput model that we presented earlier in Reference \cite{jth2010gc} can be combined with a path loss model or a database of signal strength measurements as elaborated in the considered two case studies.

While the absolute throughput is interesting in relation to performance, the impact of inaccurate and delayed information shows as a reduction in throughput performance compared to the case where information is accurate and instantly available. Therefore we consider a so-called lost throughput metric, which measures exactly this reduction in performance.

\subsubsection{Lost Throughput}
The lost throughput metric for a certain relaying policy is the difference in throughput relative to a system that has ideal location information (i.e., no positioning errors, zero delays and infinitely high update rate) and makes the optimal choices on this ideal location information. That is, it chooses the method that yields the highest throughput in each grid point. This metric is therefore useful for comparing the impact of different scenario parameter settings.
%
%
\begin{align}
    S_\text{lost}(\pi) = S_\text{ideal}(\pi_\text{opt,ideal}) - S_\text{loc}(\pi)
\end{align}
Later, in the results section, we consider the fraction of lost throughput, which is $\frac{S_\text{lost}(\pi)}{S_\text{ideal}(\pi_\text{opt})}$.
Note that there is a second way of comparing throughput, namely the difference of throughput achieved under a certain policy $\pi$ with instantaneous and accurate location information and the same policy $\pi$ with potential location errors, delay, and information loss. 
\begin{align}
    S'_\text{lost}(\pi) = S_\text{ideal}(\pi) - S_\text{loc}(\pi)
\end{align}
This latter difference can also be negative for 'poor' policies.

\section{Optimal Policy Calculation for General Scenarios with Multiple Relays}\label{sec:opt_policy_calc}

Using the Markov model and the metrics defined in the previous sections, we now show how optimal policies can be efficiently calculated with these ingredients. We thereby generalize to the scenario of $K$ relay nodes. Efficient computation here refers to polynomial time with respect to the size of the geographic region, expressed by number of states $N^2$ of the mobility model in our case. Given that there are $N^2$ geographic locations and $K$ relay candidates, each geographic location has $K+1$ potential choices, a brute-force search of enumerating and evaluating all policies would lead to exponential effort already in a mobile destination scenario; the situation even gets worse in a mobile relay scenario, as then there are $K$ mobile nodes, see further below.

As the relaying system is formulated as Markov chain model, the policy optimization can be formulated as a partially observable Markov Decision Process; however, the toolbox of algorithms for policy optimization in Markov Decision Processes \cite{puterman2009markov} are based on numericals search algorithms with risk of convergence to sub-optimal solutions. As the relaying policy in our case does not influence the transition structure of the Markov model, more efficient, exact policy optimization approaches can be utilized.  

The basic approach of the optimal policy search is:
(1) Use the Markov model with 'singleton' helper policies (i.e. policies that perform relaying in exactly one geographic coordinate) to obtain the conditional probabilities that the true coordinate of a single mobile node is $\mathpzc{x}$ conditioned on AP view is $\hat{\mathpzc{x}}$. Do this for all $K$ relays.
(2) Use then the throughput model together with the conditional probabilities from Step 1 to obtain the optimal choices for a given coordinate. 

Step (1) is the computationally more demanding one, so the resulting overall time complexity will result from Step (1) as $O(K \cdot N^2 \cdot (N^2 \cdot L)^3)$, as we will see further below. As before, $K$ is the number of mobile nodes, $N^2$ is the number of grid points, and $L$ is the number of states utilized for the information forwarding process. Step (2) will be different in cases of mobile relays as opposed to mobile destinations; consequently, the subsections \ref{sec:mc_policy_opt_mobileDest} and \ref{sec:mc_policy_opt_mobileRelay} will now differentiate these two cases. The first step however is still common to both scenarios and hence described in a general manner.

In addition to optimizing the mean throughput also the k-th percentile throughput can be optimized. Reference \cite{jrth2013vtc} analyses this metric for the scenario of a mobile destination, using a heuristic policy optimisation. A rigorous polynomial-time optimal policy search can also be defined for this metric, following the same approach as the mean throughput optimisation presented in the following.

\subsection{Calculation of conditional 
probabilities}\label{sec:mc_based_policy_optimization}

We introduce the random variable $X$ representing the true coordinate of the mobile node, and $\hat{X}$ as the random variable representing the knowledge of the AP about the MN coordinate. $(x_i)$ is the enumeration of the unique geographic coordinates, corresponding to state $i$ of the Markov mobility model.

As a first step,  we determine the probability that the true position of the mobile node is $x_j$ conditioned that the knowledge of the access point is $x_i$:
$ \Pr[X=x_j|\hat{X}=x_i]. $
This calculation forms the first step of the policy optimisation algorithm in the following way:

\subsubsection*{Step 1: Iteration}
\begin{algorithmic}[1]
\For{$i=1 \ldots N^2$}:
\State Create a singleton policy $\pi_{\text{sgl}(i)}$ which is everywhere set to 0 (D) except at coordinate $x_i$, where the policy is set to 1 (R).
\State Determine the Markov chain model (with generator $Q$)  from the  mobility model and information collection parameters; use the single entry vector $\pi_{\text{sgl}(i)}$ as the policy and calculate the steady-state solution $p$ of the Markov chain. Thereby we can obtain the desired $\Pr[X=x_j |\hat{X}=x_i]$ for the $i$th coordinate by conditioning on the AP's knowledge of the mobile node's state being 'R' in the following way:
\State $\Pr[X=x_j |\hat{X}=x_i] = \sum\limits_{s \, \in \, \text{'R' states}} p_{j,s}$
\EndFor
\end{algorithmic}
Note that propagation and throughput models are not needed in this step. 

\subsection{Calculating optimal policies in mobile destination scenario}
\label{sec:mc_policy_opt_mobileDest}
We now consider the case of a single mobile destination and $K$ static relay nodes. This scenario then requires $K+1$ throughput models $T_n(x)$, $n=0,1,...,K$, where $T_0(x)$ should reflect the direct transmission and $T_n(x)$ for $n=1,...,K$, the transmission through relay node $n$ to the mobile node at coordinate $x$.

Having obtained the conditional probabilities in Step 1, we can determine the optimal policy for a given optimisation metric. We first describe the procedure to optimise mean throughput, where we can just take the optimal decision for each coordinate locally as follows: 

\subsubsection*{Step 2 for mean throughput and mobile destination}
\begin{algorithmic}[1]
\For{$i=1 \ldots N^2$}:
  \State $\pi_\text{mean}(x_i) \coloneqq \argmax\limits_n \left\{ \sum\limits_{j=1}^{N^2}{ T_n(x_j) \cdot \Pr[X=x_j | \hat{X}=x_i]} \right\}$
\EndFor
\end{algorithmic}


Optimising percentiles of the throughput follows an analogous approach to optimising the mean throughput, though it requires a somewhat more elaborate search algorithm.


Moving to multiple mobile destinations is possible without any extra computational effort, if the stochastic mobility models of these nodes are the same and they have the same receiver (so that the throughput model is also the same across all nodes). In that case, the access point just applies the decision which it obtains from the relay policy that it has calculated for one of the nodes. If the destination nodes show different stochastic mobility patterns or different receivers, optimal policies for each of the mobility patterns need to be obtained; this is doable with linear effort increase. Note however, that also delay and loss parameters for the location updates may need to be adapted for an increasing number of nodes.

\subsection{Calculating optimal policies in mobile relay scenario}
\label{sec:mc_policy_opt_mobileRelay}
In the case of $K$ mobile relays, calculating the whole policy would require a table with $N^{2K}$ entries of value range $0,1,...,K$, representing the relay node choice in case of a specific coordinate allocation of the $K$ relay nodes. To avoid this large table representation, an efficient representation can make use of the fact that the resulting throughput through relay node $r$ is independent of the other relay node coordinates; so it would be sufficient to create  $K+1$ tables of size $N^2$, in which the resulting average throughput for a specific relay $r$ at assumed position $x_1,...,x_{N^2}$ is calculated. During run-time, the AP would then need to search for the maximum of the $K+1$ values from the different tables and pick the relay node (or direct transmission) that maximizes the throughput.

\subsubsection*{Step 2' for mean throughput and mobile relays}\mbox{}
\begin{algorithmic}[1]
\For{$i=1 \ldots N^2$}:
\For{$r=0 \ldots K$}:
\State $\Gamma_r(i) \coloneqq \sum\limits_{j=1}^{N^2}{T_n(x_j) \cdot \Pr[X=x_j | \hat{X}=x_i]}$
\EndFor
\EndFor
\end{algorithmic}
where $\Gamma_r(i)$ is a throughput table for relay $r$ at coordinate $x_i$.


Note that the system definition which is basis for the policy optimisation is not using any location prediction. Location prediction would not even be needed to be algorithmically included explicitly, but it would instead be sufficient to look at relay policies that can use the current assumed coordinate and the previously assumed coordinate as input. The Markov model could be extended to include some memory on previous coordinates for such more powerful policies. Such extensions are for future study.

\section{Case Study 1: Outdoor Open Field}\label{sec:case_study_1}
For demonstrating the application of the proposed model, we consider first an ideal case study that reflects the scenario in Fig. \ref{fig:3-node-system}. Access point AP and destination node D are static, whereas the relay node R moves according to the Markov mobility model presented in Section \ref{sec:markov_mobility}. The scenario assumes IEEE 802.11a communication with a relay-enabled MAC-layer as mentioned in Reference \cite{jth2010wcnc}. Table \ref{tab:default_scenario_parameters} lists the used scenario and simulation parameters.

The value of the network delay rate, $\mu$, has been calculated using the formulas from Reference \cite{jun2003theoretical}, given an assumed frame payload of 28 bytes for location coordinate and node identifier as specified in Reference \cite{jth2010wcnc}.

\begin{table}
    \caption{Default scenario and simulation parameters.}
    \label{tab:default_scenario_parameters}
    \centering
    \scriptsize
        \begin{tabular}{|l|c|}
        \hline
        \textbf{Parameter}          & {\textbf{Value}} \\ \hline
        Dimensions              & {$80 \times 80$ m$^2$} \\ \hline
        No. of grid points          & {$10 \times 10$} \\ \hline
    Grid spacing $d$            & {$8 \times 8 $m$^2$} \\ \hline
    AP coordinate & {$(16,40)$} \\ \hline
    Destination coordinate      & {$(64,40)$} \\ \hline
        Data transmission interval  & {$25$ s} \\ \hline
        Measurement delay rate $\mu$ & {$1/(0.3748 \cdot 10^{-3})$ s$^{-1}$}  \\ \hline
        Wireless link error probability $p_\text{loss}$ & {0} \\ \hline
    Ricean $K$              & {$6$} \\ \hline
    Path loss exponent $n$ & {$2.9$} \\ \hline
    Data frame payload $B_\text{MSDU}$          & {$1500$ bytes} \\ \hline\hline
    \textbf{Scenario A: low dynamics} & \\ \hline
        Measurement update rate $\tau$                  & $1/5$ s$^{-1}$\\ \hline
        Location error std. dev. $\sigma_\text{loc-err}$ & 0 m \\ \hline
    \textbf{Scenario B: high dynamics/challenging} & \\ \hline
        Average movement speed $\bar{v}$  & $5$ m/s \\ \hline
        Measurement update rate $\tau$    & $1/25$ s$^{-1}$\\ \hline
  Location error std. dev. $\sigma_\text{loc-err}$ & 5 m \\ \hline
        \end{tabular}
\end{table}

%

\subsection{Throughput Model}\label{sec:tp_model}
In order to calculate the expected throughput for the 802.11a based case study, we use first a standard log-distance path loss model (e.g., \cite{durgin1998mam}) to determine the path loss on the direct and two relay links, given the distances between the nodes for different relay positions:
\begin{equation}
PL = PL_{d_0} + 10 n \log_{10} (d/d_0) ~ \text{[dB]} \label{eq:pl_model}
\end{equation}
where $PL_{d_0}$ is the path loss at a reference distance $d_0$, $n$ is the path loss exponent, and $d$ is the link distance.

Based on the path loss and the scenario parameters given in Table \ref{tab:default_scenario_parameters}, we use the throughput model we have developed previously in Reference \cite{jth2010gc} and slightly updated in Reference \cite{nielsen2011location}. 
Throughput is given by the ratio of expected delivered data divided by expected transmission time. By using bit-error-rate models to calculate the frame error probabilities (see \cite{jth2010gc,nielsen2011location} for details), the throughput for the direct link can be calculated as:
\begin{align}\label{eq:tp_model:S}
S_\text{dir} = \frac
{P_\text{suc} \cdot B_\text{MSDU}}
{E[T_\text{tx}]}
\end{align}
where $P_\text{suc}$ is the probability of a successful MAC layer frame delivery, $E[T_\text{tx}]$ is the duration of a MAC frame delivery attempt, and $B_\text{MSDU}$ is the MAC payload size given in octets.
In the following, we use the indices 1 and 2 to indicate the AP-R and R-D transmissions. The throughput for the two-hop relaying algorithm is calculated as:
\begin{align}\label{eq:clo:S_rel}
S_\text{rel} = \frac
{(P_\text{suc}^\text{pri,1} P_\text{suc}^\text{pri,2} + P_\text{suc}^\text{sec,1} P_\text{suc}^\text{sec,2}) \cdot B_\text{MSDU}}
{E[T_\text{tx}^\text{pri,1}] + E[T_\text{tx}^\text{pri,2}] + E[T_\text{tx}^\text{sec,1}] + E[T_\text{tx}^\text{sec,2}]} .
\end{align}

\begin{figure}[t]
\centering
    \subfigure[Relayed transmission]{
        \includegraphics[height=4.5cm]{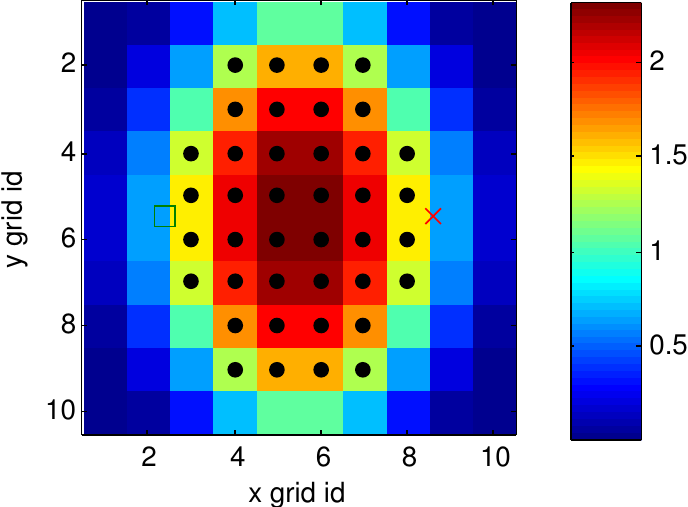}
        \label{fig:relay_map_relayed}
    }
    \subfigure[Direct transmission]{
        \includegraphics[height=4.5cm]{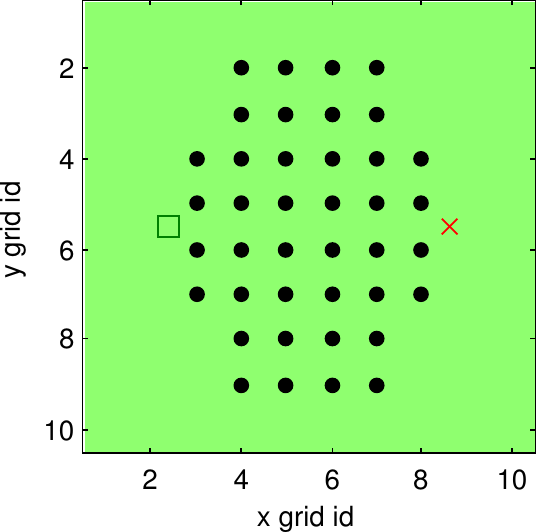}
        \label{fig:relay_map_direct}
    }
\caption{Throughput [Mbit/s] per grid point. The black dots show where the relayed throughput is higher than direct throughput; this is denoted the \textit{standard policy}. The green square is the AP and the red cross is the destination.}
\label{fig:relay_map}
\end{figure}

The throughput model is used in this work to estimate the transmission throughput functions $T_\text{D}(m)$ and $T_\text{R}(x_i)$ for each of the $N^2$ grid points $x_i$. Fig. \ref{fig:relay_map} shows the throughput for a $10 \times 10$ grid realisation of the Case Study 1 scenario.

Notice that the path loss, bit error and throughput models presented above can be exchanged to fit other scenarios or systems for which relaying is considered. The simple models above are however advantageous for a first case study, as they use a small number of parameters.


\subsection{Location Error}
It is assumed that it is possible to obtain a description of the location error from the used localisation system. Ideally, this would be a likelihood distribution function of the considered area, for example based on the outcome of a particle filter localisation algorithm. Often, only a measure of the variance is provided, and in such cases it is necessary to make some assumptions on the error. Commonly and in this work for simplicity, it is assumed that the location error of the positioning system follows a symmetric, truncated multi-dimensional Normal distribution: $\tilde{\mathpzc{x}}(t)=\mathpzc{x}(t) + N(\mathbf{\mu}_\text{err},\mathbf{\sum}_\text{err})$, where $\mathbf{\mu}_\text{err}$ is the bias introduced by the position system and $\mathbf{\sum}_\text{err}$ is the covariance matrix. In our examples, we assume that $\mathbf{\mu}_\text{err}=0$ , and the covariance matrix is a diagonal matrix with entries set to $\sigma_\text{loc-err}$, the latter called the location error standard deviation.
However, the probability mass that falls outside the considered geographic region is cut away. This corresponds to a location system in which an invalid location estimate is discarded. Specifically the location error probability distribution matrix is built in the following way:
\begin{enumerate}
\item Let $x_i$ be the center coordinate of the $i$-th grid point corresponding to the relay's true position. 
\item For this $i$th grid point, draw the probability density value of each grid center point $x_j$ from the Gaussian 2-dimensional location error function for the given value of $\sigma_\text{loc-err}$, and store these probabilities in the element $E_{i,j}$ of the error matrix $E$. Note that $E$ has dimensions $N^2 x N^2$ when the Markov mobility model covers $N^2$ geographic states.
\item Re-normalise the rows of $E$, so that the sum of all entries in a row is 1. 
\end{enumerate}

\subsection{Parametric Studies and Simulation Validation}
For model validation and performance evaluation, this work uses the parameters listed in Table \ref{tab:default_scenario_parameters} with the proposed Markov chain model and a simulation model. The simulation model is Matlab based and implements the system model presented in Section \ref{sec:system_description}. Previous work in Reference \cite{jrth2012wcnc} presented results for a subset of the parameters studied in the following.
For both the Markov chain model and simulation model, the considered mobility model grid size is 10x10, which was determined heuristically from comparisons to continuous mobility simulations in Reference \cite{jrth2012wcnc}.

In all simulation results, the so-called standard policy (std. policy) is determined by comparing the expected throughput of direct or relayed transmissions for the believed relay position and then use the mode whose throughput is highest.
This \emph{std. policy} is exemplified by the dots in Fig. \ref{fig:relay_map} and it is the only policy that is used for the simulation results. Here it serves to validate the corresponding Markov chain curve \emph{MC model - std. policy}. For the MC model, also the location independent policies of always transmitting directly or relayed are shown (\emph{direct} or \emph{relaying}). The \emph{inv. policy} is simply the inverse of the standard policy, providing a poor choice as reference for comparison.
Finally, the curve named \textit{opt. policy}, uses the optimisation algorithm from Section \ref{sec:opt_policy_calc} to determine the optimal policy, with \textit{average throughput} as the optimisation criterion.

Results are in the following presented for two different Scenarios A (low dynamics) and B (high dynamics), as defined in Table \ref{tab:default_scenario_parameters}.

\begin{figure*}
\centering
    \subfigure[Low dynamics scenario A]{
    \includegraphics[width=7cm]{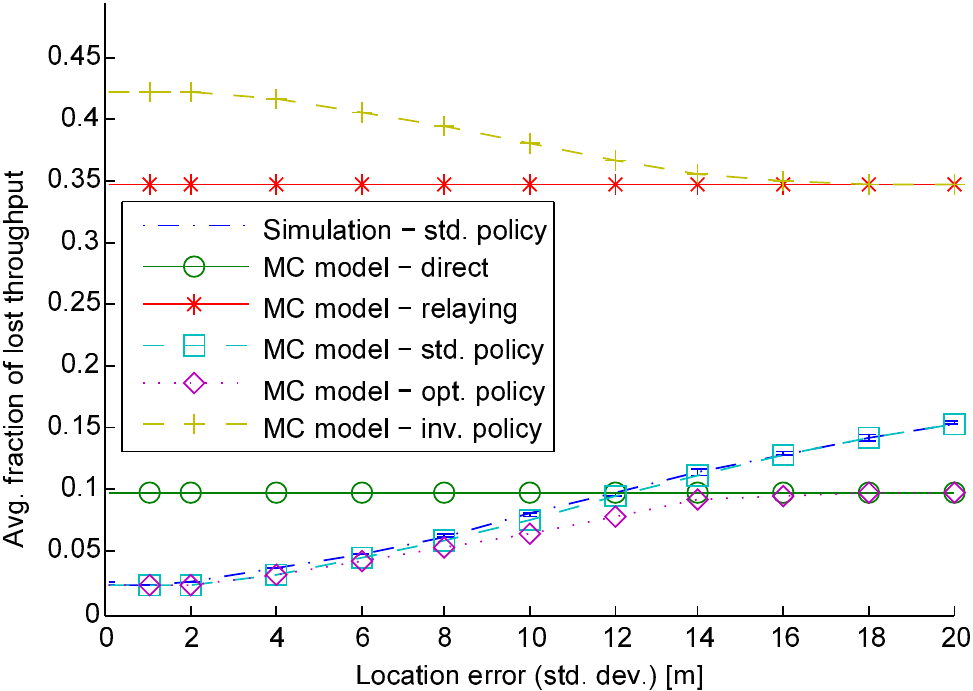}
    \label{fig:mmPr-loc-err_lost_throughput}
    }
    \subfigure[High dynamics scenario B]{
   \includegraphics[width=7cm]{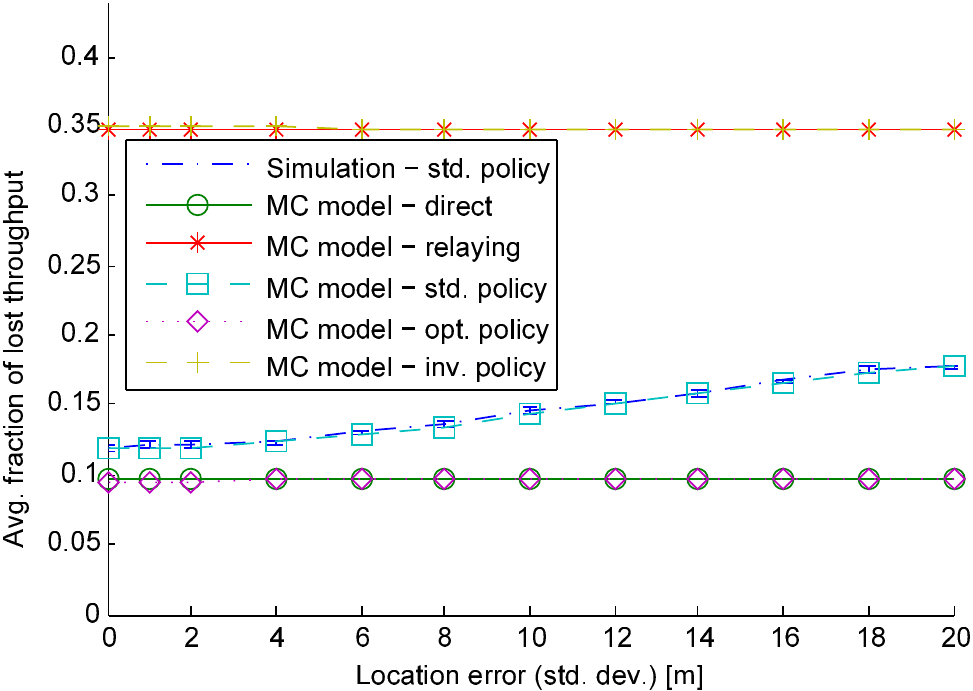}
    \label{fig:mmPr-loc-err-fast_lost_throughput}
    }
   \subfigure[Fraction of positions where relaying is preferred for optimal policy]{
    \includegraphics[width=7cm]{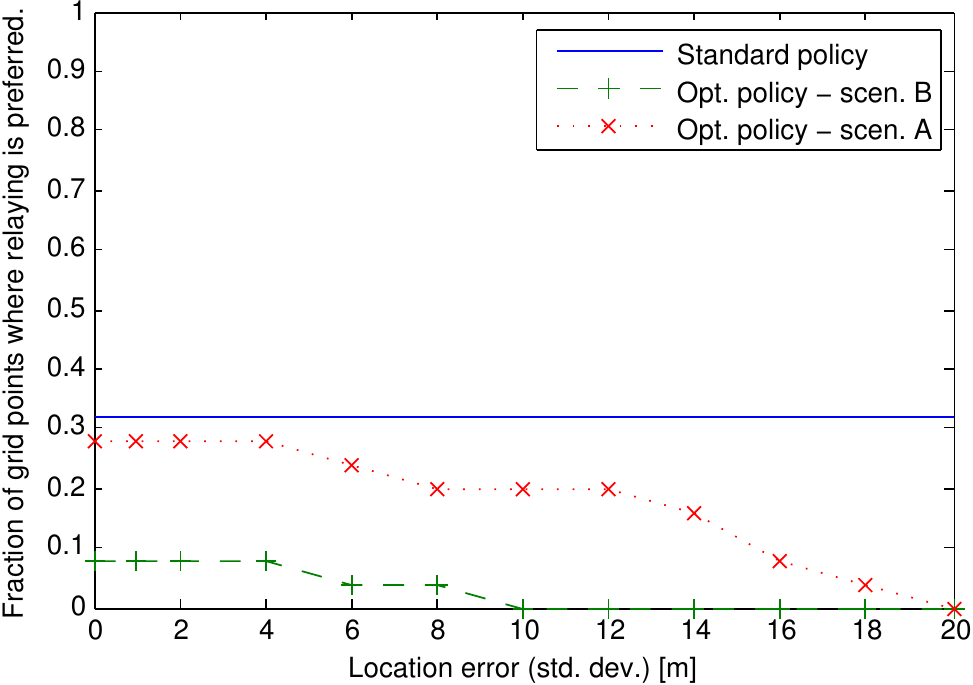}
\label{fig:mmPr-loc-err-both_R_count}
   }
\caption{Lost throughput and fraction of positions where relaying is preferred for varying location error.}
\label{fig:results_loc-err}
\end{figure*}

The plots in Fig. \ref{fig:results_loc-err} show the impact of adding a random 2D Normally distributed location error to the coordinate of the mobile relay's location that is sent periodically from the mobile relay to the AP. This corresponds to the error that would arise with an actual location system based on for example GPS or indoor radio localisation.

Besides the accurate resemblance of the simulation and Markov chain results for the standard policy, the most noteworthy result here is for the optimal policy in Fig. \ref{fig:mmPr-loc-err_lost_throughput}. This shows that policy optimisation can achieve a substantial reduction of lost throughput, especially when the location error increases above 12 m, which could occur in harsh indoor environments with many metallic objects that make the location estimation difficult. Here, the $33\%$ reduction in lost throughput from $0.15$ to $0.10$ (at 20 m) demonstrates the benefit of policy optimisation. In absolute terms, this improvement amounts to around $0.1$Mbit/s; from 1.4 to 1.5 Mbit/s, where the avg. throughput with exact location information and no access delays is 1.65 Mbit/s. The plot in Fig. \ref{fig:mmPr-loc-err-both_R_count} shows that the policy optimisation achieves a gain by reducing the number of points in which relaying is used.


%

The plots for the high dynamics scenario in Fig. \ref{fig:mmPr-loc-err-fast_lost_throughput} show similar results, but here the possible reduction in lost throughput is from $0.18$ to $0.10$, i.e., a reduction of almost $45\%$.
The corresponding curve for the high dynamics scenario in Fig. \ref{fig:results_loc-err} shows that location information becomes less useful with increasing dynamics, so that relay decisions based on it are taken in fewer places.

Overall we can conclude that policy optimisation is necessary to ensure that location based relay selection is at least as good as a fixed policy (always direct or always relay) when location error increases. In the latter case, the non-optimised location-based relay selection can turn out worse than a fixed policy.

\begin{figure*}
\centering
    \subfigure[Low dynamics scenario A]{
    \includegraphics[width=7cm]{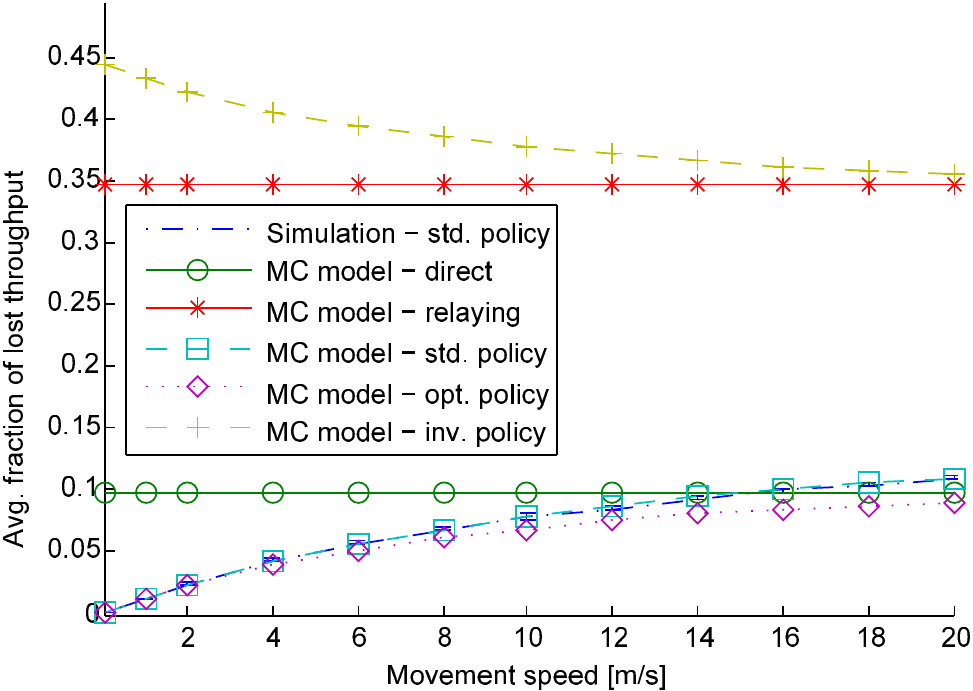}
    \label{fig:mmPr-speed_lost_throughput}
    }
    \subfigure[High dynamics scenario B]{
    \includegraphics[width=7cm]{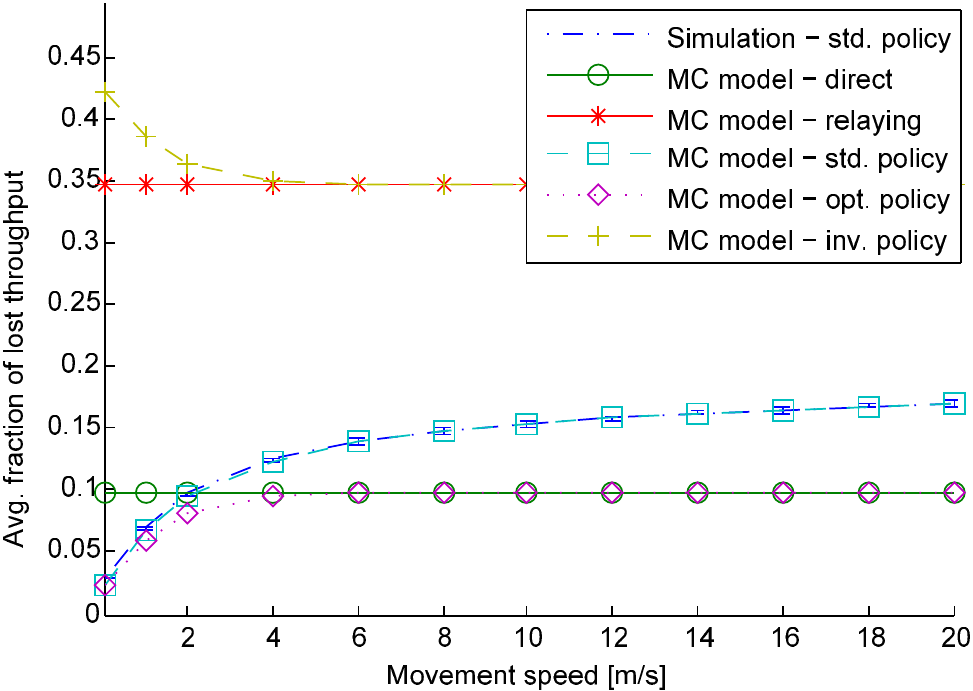}
    \label{fig:mmPr-speed-fast_lost_throughput}
    }
   \subfigure[Fraction of positions where relaying is preferred for optimal policy]{
    \includegraphics[width=7cm]{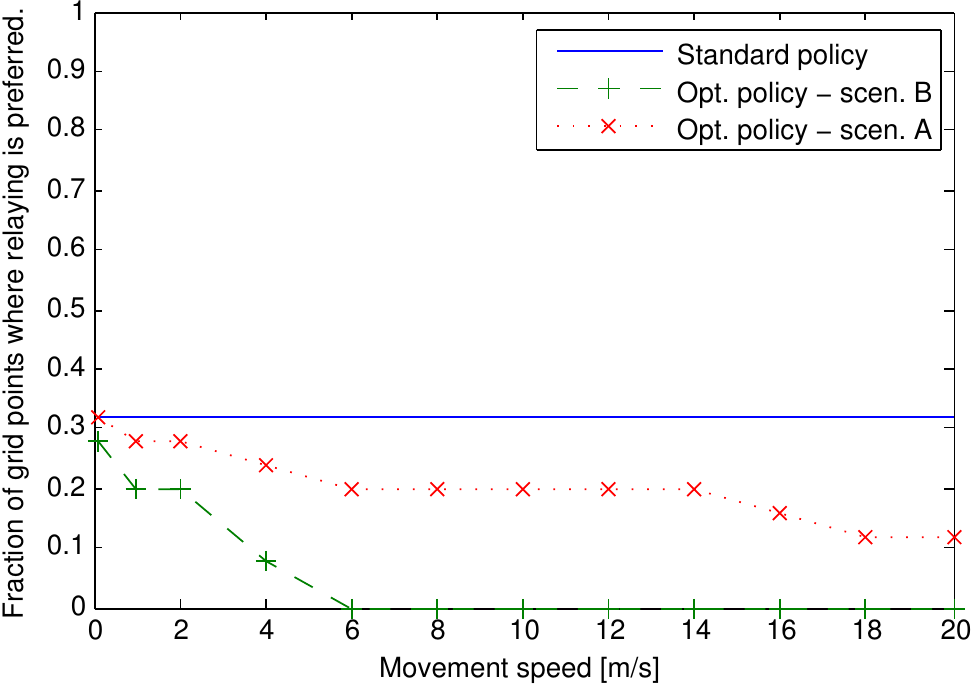}
\label{fig:mmPr-speed-both_R_count}
   }
\caption{Lost throughput and fraction of positions where relaying is preferred for varying movement speed.}
\label{fig:results_speed}
\end{figure*}

Fig. \ref{fig:results_speed} shows results for varying the movement speed. The nearly identical curves for the standard policy with simulation and MC model in Fig. \ref{fig:mmPr-speed_lost_throughput} show that the impact of movement speed is also accurately accounted for in the model. Further, since the optimal policy brings only a modest gain of $0.02$ at $20$m/s we can conclude that the standard policy (black dots in Fig. \ref{fig:relay_map}) is sufficient in this scenario. Fig. \ref{fig:mmPr-speed-both_R_count} shows the number of grid points in which relaying is preferred for the determined optimal policy and we see that direct transmissions are preferred more often as the movement speed increases. This result corresponds with the intuition that a wrongly chosen relay transmission may be more expensive than a wrong direct transmission (if a direct transmission is possible), since the direct transmission quality is guaranteed in the considered scenario where the AP and destination nodes are static, whereas a wrongly chosen relay transmission may give close to zero throughput, c.f. Fig. \ref{fig:relay_map}. 

In Fig. \ref{fig:mmPr-speed-fast_lost_throughput} we show similar results, however for the considered high dynamics scenario with longer measurement update intervals. The optimised policy brings a slight improvement compared to the standard policy, which is achieved by further limiting the number of relay transmissions as shown in Fig. \ref{fig:mmPr-speed-both_R_count}.

Decreasing the rate of location update messages turns out to show a similar impact as increasing the movement speed of the mobile node. Corresponding results are not shown here for space reasons. 



\subsection{Policy choices}
\begin{figure}
  \centering
  \includegraphics[width=7cm]{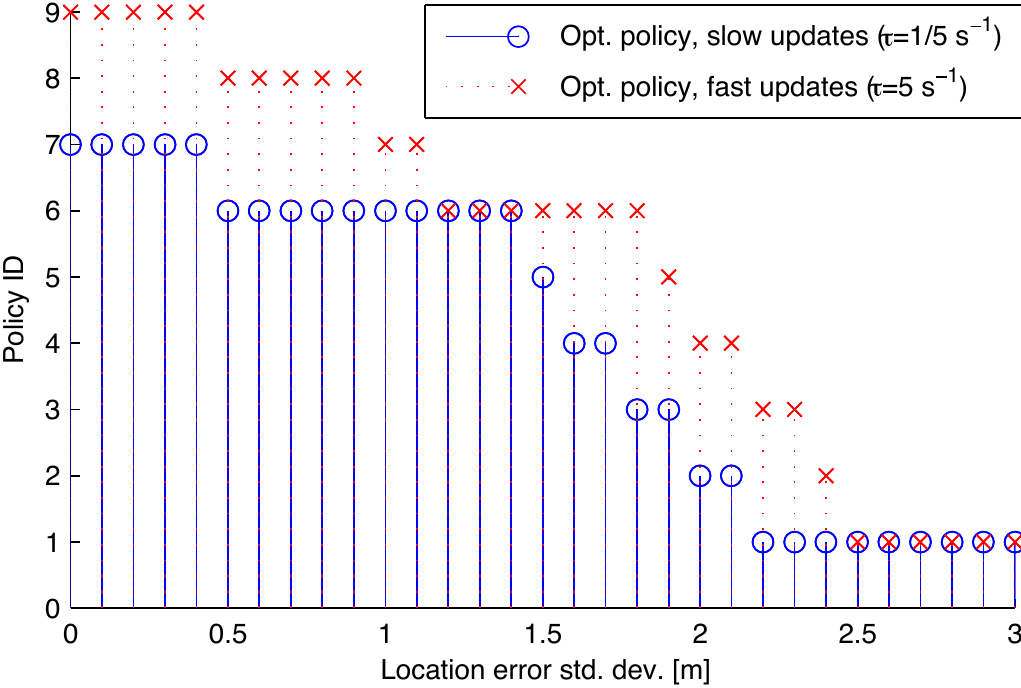}
  \caption{Choices in optimised relay policies for increasing location error. The scenario is based on Scenario B ($v=5m/s$), using the specified $\tau$ and location error values. The policy with ID=1 is all-zero, i.e., relaying is never used.}
  \label{fig:mmPr_loc_err_policy_stem}
\end{figure}

Fig. \ref{fig:mmPr_loc_err_policy_stem} shows the change in choices of optimised policies when increasing the location error for two different values of $\tau$, corresponding to a slow and a fast update of location information. Each unique policy has been assigned an ID, which is used on the y-axis of the plot. From the plot it is clear that the policies with IDs 1-7 are used for both values of $\tau$, however only for the fast updates the policies with IDs 8 and 9 are used. In this case, when the update rate is fast, the policy optimisation is better able to mitigate the effect of mobility than with a slow update rate. However, as the location error increases, this becomes the dominant factor and at 2.5 m location error, the optimal policy is in both cases to use only direct transmissions.

\subsection{Model-based adaptive update rate}
\begin{figure}
  \centering
  \includegraphics[width=7cm]{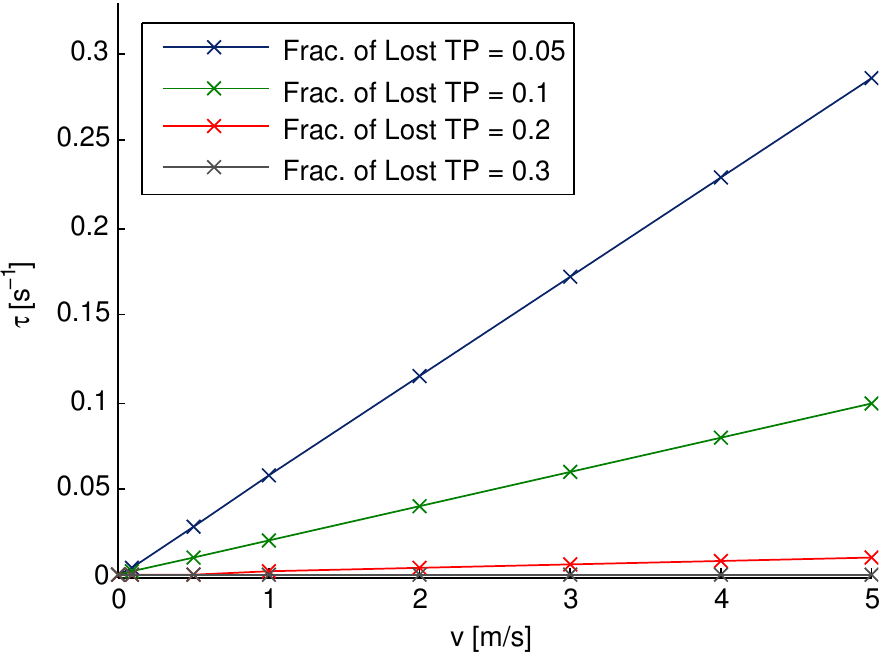}
  \caption{Required $\tau$ for different mobility speeds to achieve certain levels of lost throughput, assuming no location error.}
  \label{fig:mmPr_tau_v_lost-tp_plot}
\end{figure}

A possible application of the proposed model is to use it for adjusting parameters such as the measurement update frequency according to the scenario conditions. Since measurement updates generate signalling overhead, it is desirable to be able to determine the update rate to achieve a required level of performance. In Fig. \ref{fig:mmPr_tau_v_lost-tp_plot} we have used the MC model to calculate the update rate $\tau$ that is required to achieve a certain level of lost throughput, for varying movement speeds, assuming zero location error.
The figure shows that this required update rate $\tau$ is approximately linearly dependent on the average movement speed, however the slope depends on how much lost throughput can be tolerated. 
In principle, the transmission delay $\mu$ makes the curves not completely linear, however, since $\mu$ is very small compared to $\tau$ the curves visually appear linear.

\subsection{How accurate should localisation be?}
By also taking into account the location error, a similar analysis can be conducted to answer the question: ''How accurate should localisation be for location based relay selection to be beneficial?''. It is necessary to specify a precise condition that must be fulfilled to answer this question. As an example, we specify this condition as being able to achieve at least a certain percentage of additional throughput than with a non-location based direct or relayed relaying policy. Specifically, we use:
\begin{align}
\gamma_\text{benefit} < \frac{S_\text{loc}}{\max(S_\text{dir},S_\text{rel})}
\end{align}
where $\gamma_\text{benefit}>1$ is the benefit threshold, $\max(S_\text{dir},S_\text{rel})$ is the maximum achievable throughput with a constant relaying policy, and $S_\text{loc}$ is the achievable throughput with a location based policy. For the latter, we will consider both the standard policy and the throughput optimised policy.
For this analysis, we use the Markov chain model to iteratively search for the $\tau$-value that gives us exactly $\gamma_\text{benefit}$ benefit, given the scenario outlined earlier in this section, for a few selected movement speeds with the standard and optimised policies.

\begin{figure}
  \centering
  \includegraphics[width=7cm]{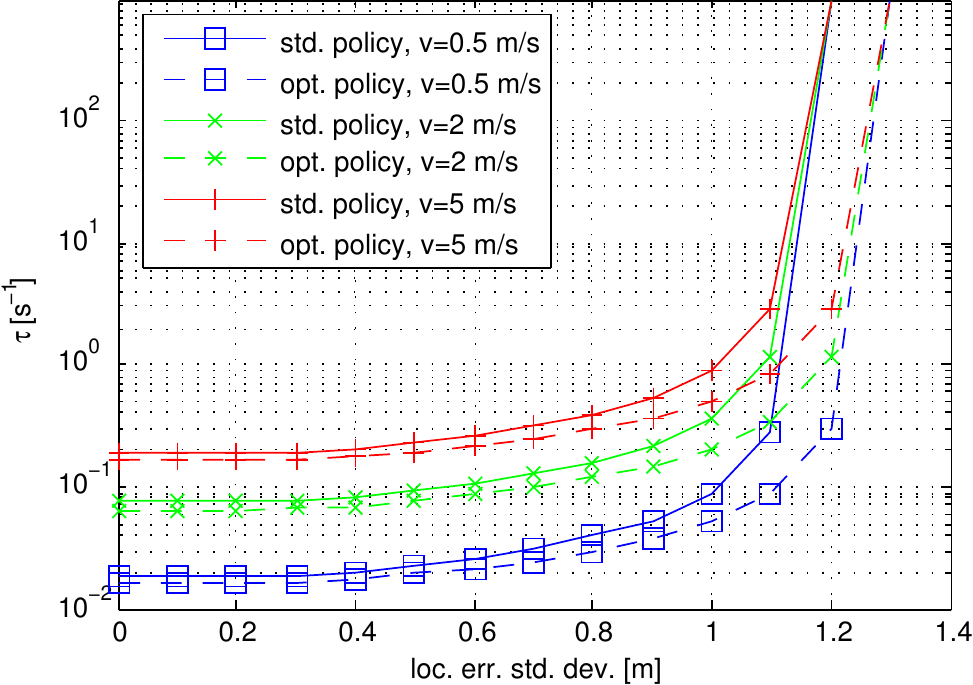}
  \caption{Required $\tau$ for different mobility speeds and policies to achieve $5\%$ of throughput benefit.}
  \label{fig:mmPr_tau_loc_err_plot_5}
\end{figure}

The plot in Fig. \ref{fig:mmPr_tau_loc_err_plot_5} shows the obtained values of $\tau$ for a benefit of $5\%$, i.e., $\gamma_\text{benefit}=1.05$. Notice that $\tau$-values larger than 1000 were not considered, as sub-millisecond sample rates are unlikely in a practical system.

%
%

It can be seen in this plot that for the given scenario and the specified benefit conditions, the maximum acceptable location error is around $1.1-1.2\,$m, depending on the applied policy. Furthermore, it is clear that there is a highly non-linear relationship between the given location error and the required $\tau$-value, where up to a certain level of location error, the required $\tau$-value does not increase very much, but beyond this level of location error, it grows with a vertical asymptote. Notice that the optimal policy in Fig. \ref{fig:mmPr_tau_loc_err_plot_5} is able to shift the asymptote of this rapid increase, meaning that a slightly higher level of location error can be tolerated with the optimised policy compared to the standard policy. The sudden rapid increase of these curves is contrary to the results in Fig. \ref{fig:mmPr_tau_v_lost-tp_plot}, where the update rate $\tau$ can directly counteract the impact of increasing mobility speed. However, the negative impact of location error cannot be similarly mitigated by increasing the update rate $\tau$. What actually happens in Fig. \ref{fig:mmPr_tau_loc_err_plot_5} is that the update rate $\tau$ is used to mitigate the mobility induced errors so that the combined mobility error and location error allows the benefit condition to be fulfilled. 
When investigating Fig. \ref{fig:mmPr_tau_loc_err_plot_5} closely, the horizontal distance between the two asymptotes is approximately $0.1\,$m, so the policy optimisation can mitigate an additional location error of that size.

Even though the localisation accuracy is in this analysis treated as a scenario parameter that is a property of the used localisation system, it can sometimes be considered as a  controllable parameter. Specifically, the location accuracy can be improved or decreased by using more or fewer measurements and/or more or less advanced location estimation algorithms. If less location accuracy can be tolerated, it is possible to extend battery life-time of the mobile devices and reduce signalling overhead.
One example is given in Reference \cite{amiot2012evaluation}, where the authors investigate the trade-off of using different types of ranging measurements and algorithms of different complexity. A location based relay selection algorithm should therefore not necessarily be considered as an independent system block, as it could most likely benefit from being able to increase location accuracy on demand by using additional measurements and processing resources, when policy optimisation alone is not sufficient to meet the desired performance goal. On the other hand, it would also be possible to use fewer measurements and processing resources on the location estimation, in cases where a slight increase in update rate could allow the desired performance goal to be reached.

\section{Case Study 2: Measurement Based Indoor}\label{sec:measurement_case}
For demonstrating the application of the proposed model in a more complex scenario, we consider a case study that reflects an indoor scenario in Fig. \ref{fig:layout}. The AP and D nodes are static, whereas the R node moves according to the Markov mobility model shown in Fig. \ref{fig:mobility_model_in_room_layout}. The nodes are assumed to use 802.11g based radios, with a relay-enabled MAC-layer as mentioned in Reference \cite{jth2010wcnc}. Notice that this case study assumes 802.11g and not 802.11a as in the first case study. Furthermore, since the geographic area is relatively small compared to the typical range of 802.11g, the transmission parameters have been scaled down to imitate a scenario in which relaying is usable. Table \ref{tab:default_scenario_parameters} lists the used scenario and simulation parameters. The specific scenario is described in the following.

\subsection{Ray Tracing Simulation}
\begin{figure}[t]
  \begin{center}
    \includegraphics[width=0.6\textwidth]{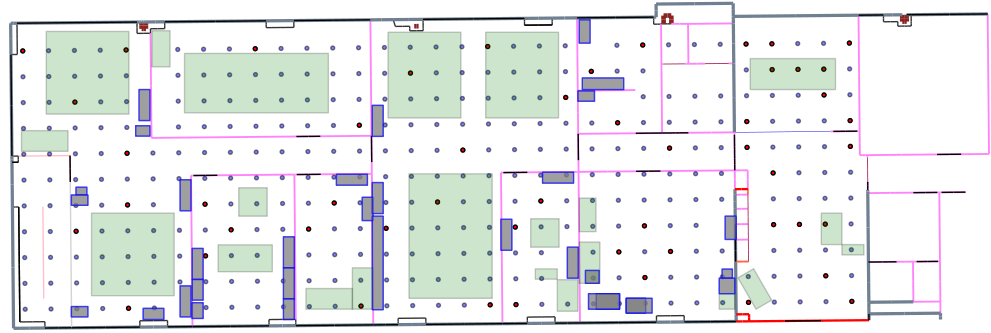}
  \end{center}
  \caption{The indoor Layout with furniture and the used grid of points (pseudo AP and destination nodes are in red.)}
  \label{fig:layout}
\end{figure}

A ray tracing simulator named PyLayers \cite{pylayers} has been used to produce a large set of realistic received signals on a uniform grid of 51 pseudo Access Points $\times$ 363 Mobile Stations (MSs), which covers the office building described in \cite{WD41}. Each grid point is $1~\text{m} \times 1~\text{m}$.
In this paper, only a subset covering the first 15 of 51 pseudo AP positions and the first 147 of the 363 MS positions is used, corresponding to the leftmost 2/5 of the building shown in Fig. \ref{fig:layout}.

For our simulations the IEEE 802.11g channel 1 was assumed, corresponding to a center frequency of $2.412\,$GHz and channel bandwidth of $20\,$MHz.
For a more extensive description of the ray tracing for this scenario, see Reference \cite{amoit2013pylayers}.

\subsection{Used Throughput Model}
Given the extracted path loss for a narrow band corresponding to IEEE 802.11g channel 1, we estimate the achievable throughput from the path loss using the throughput model described in Section \ref{sec:tp_model}.


%
\begin{figure}
\centering
        \includegraphics[height=6cm]{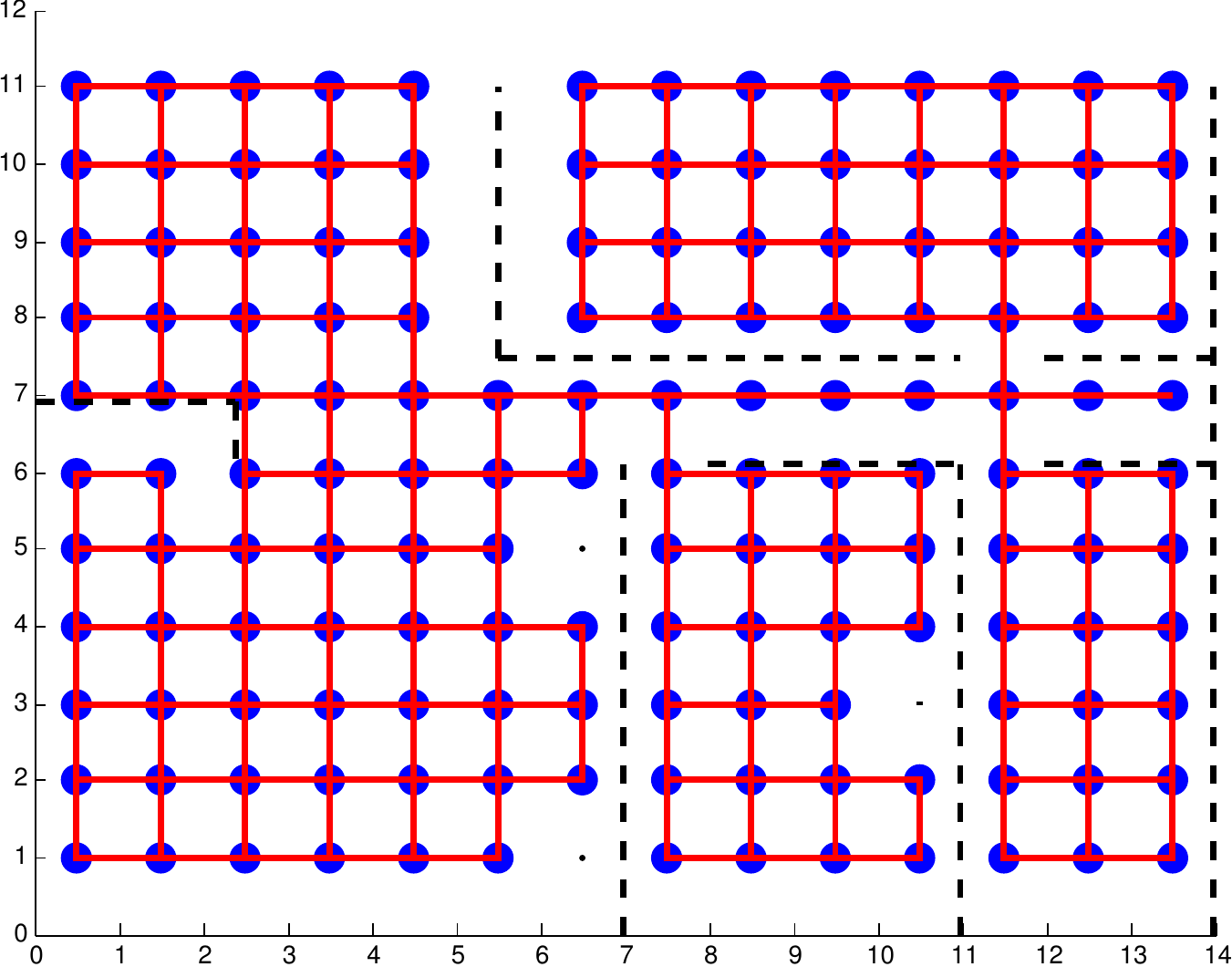}
    \caption{Mobility model in room layout. Red lines are possible movements between states (blue dots) and black dashed lines are walls.}
\label{fig:mobility_model_in_room_layout}
\end{figure}


%



\begin{figure}
\centering
    \includegraphics[height=6cm]{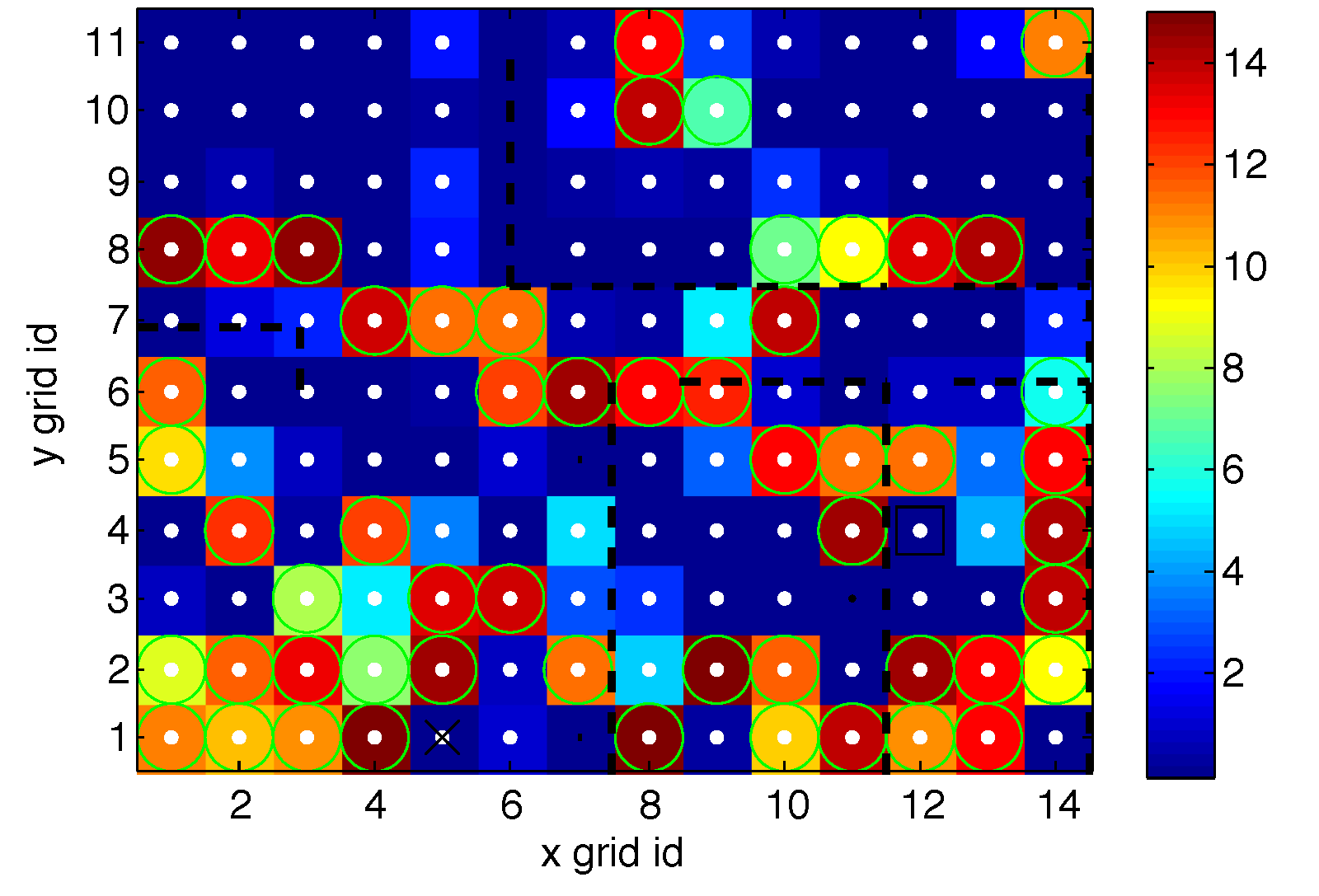}
    \caption{AP and D positions are marked by a black square $(12,4)$ and black cross $(5,1)$, respectively. The colour of each grid point shows the achievable relay throughput [Mbit/s], if the mobile relay is in that position. The green circles mark the grid points in which the relay throughput exceeds the direct throughput.}
\label{fig:throughput_model_in_room_layout}
\end{figure}

\subsection{Candidate Policies}
For evaluation we define first a reference (Ideal) policy that assumes instantaneous collection and perfect information and that the best possible choice direct/relay is always made.
Besides the ideal, we consider three location based policies, a heuristic one which relies on a coarse room-level localisation accuracy (Heuristic) and two others that both require grid-level accuracy; one uses the locally optimal standard policy (Locally Optimal), i.e., which mode has the highest expected throughput in the currently believed position, and another (Optimised Policy), which furthermore uses the policy optimisation algorithm described in Section \ref{sec:mc_based_policy_optimization}. For comparison we consider also the two fixed policies of always transmitting directly or always using the relay.
Since the information collection has an impact on the performance when using the location-based policies, the three latter policies will be analysed considering delayed information collection.
A summary is given in the table below:

\begin{center}
{\scriptsize
\begin{tabular}{|l|c|c|}
\hline
\textbf{Name} & \textbf{Loc. accuracy} & \textbf{Info. collection}\\ \hline
Ideal & grid level & instantaneous \\ \hline
Optimised policy & grid level & delayed \\ \hline
Locally optimal & grid level & delayed \\ \hline
Heuristic & room level & delayed \\ \hline
Always direct & - & - \\ \hline
Always relay & - & - \\ \hline
\end{tabular}
}
\end{center}

For the heuristic scheme we have defined that the relay will only be used when within the two rooms delimited by the rectangle between the two corners (8;1) to (11;6).

\begin{table}[t]
    \caption{Default scenario parameters for use case study.}
    \label{tab:default_use-case_2_parameters}
    \centering
    \scriptsize
        \begin{tabular}{|l|c|}
        \hline
        \textbf{Parameter}          & {\textbf{Value}} \\ \hline
        Measurement delay rate $\mu$ & {$10^5$ s$^{-1}$} \\ \hline
        Network loss probability $p_\text{loss}$ & {0} \\ \hline
    Noise floor                 & {$-85$ dBm}\\ \hline
    Ricean $K$              & {$6$} \\ \hline
    Data frame payload $B_\text{MSDU}$  & {$1500$ bytes} \\ \hline
        Average movement speed $\bar{v}$ & $0.5$ m/s\\ \hline
        Measurement update rate $\tau$  & $1$ s$^{-1}$ \\ \hline
        \end{tabular}
\end{table}

\subsection{Results and Discussion}
\begin{figure}[t]
\centering
    \subfigure[Location error]{
        \includegraphics[width=0.465\textwidth]{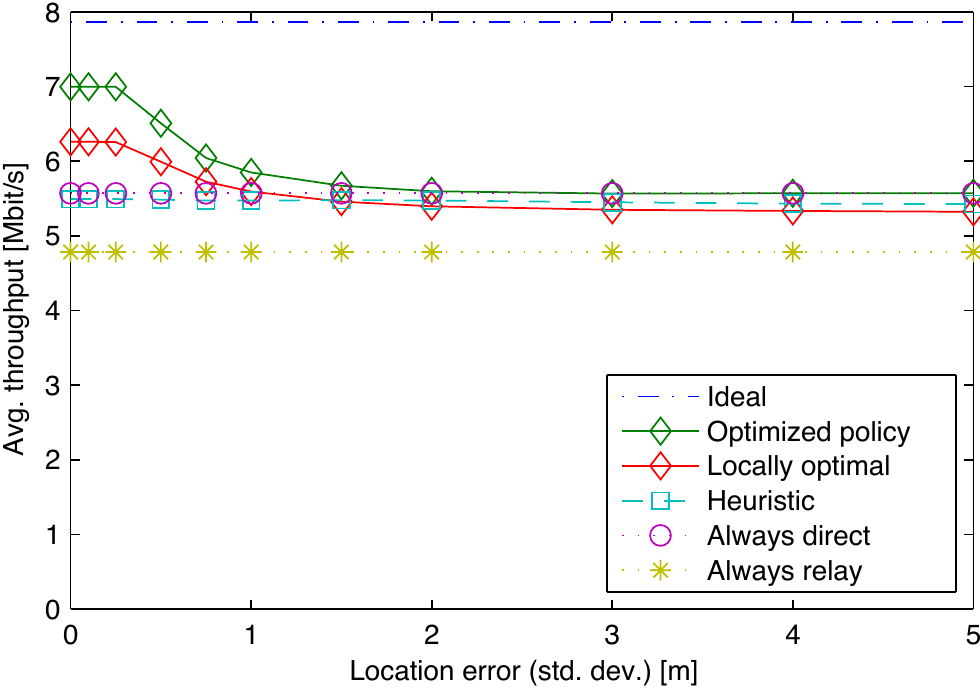}
    \label{fig:results_ray_loc_err}
    }
    \subfigure[Relay movement speed]{
        \includegraphics[width=0.465\textwidth]{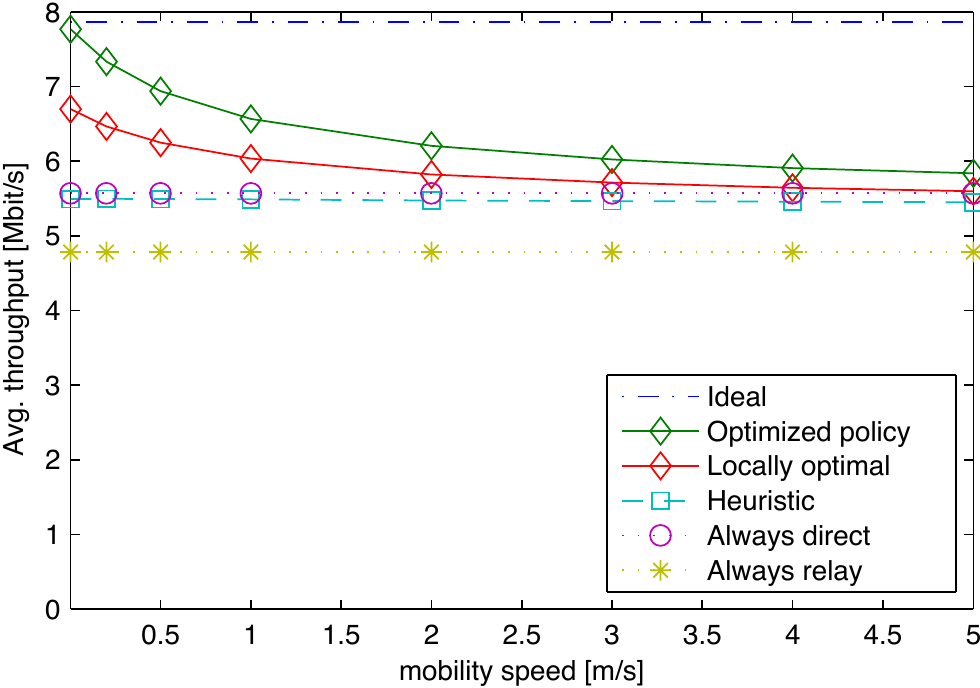}
    \label{fig:results_ray_speed}
    }
    \caption{Average throughput for varying location error and relay speed.}
\label{fig:results_case_2}
\end{figure}

For evaluating the performance of the different schemes described in the section above, we have applied the constrained mobility model shown in Fig. \ref{fig:mobility_model_in_room_layout}, the two throughput models shown in Fig. \ref{fig:throughput_model_in_room_layout} and the default parameters listed in Table \ref{tab:default_use-case_2_parameters} on the Markov chain model described in Section \ref{sec:mc_model}. The results obtained when varying the location error and relay movement speed are presented in Fig. \ref{fig:results_case_2}.
%
%
%


For both parameters being varied with the optimised policy and locally optimal schemes, the results show that optimised relaying does provide a significant benefit of more than 20\% compared to using a static relay strategy (always direct or always relay). The gain compared to using the locally optimal strategy is around 10\%. 

%
%

%
\begin{figure*}[t]
\centering
    \includegraphics[width=0.8\textwidth]{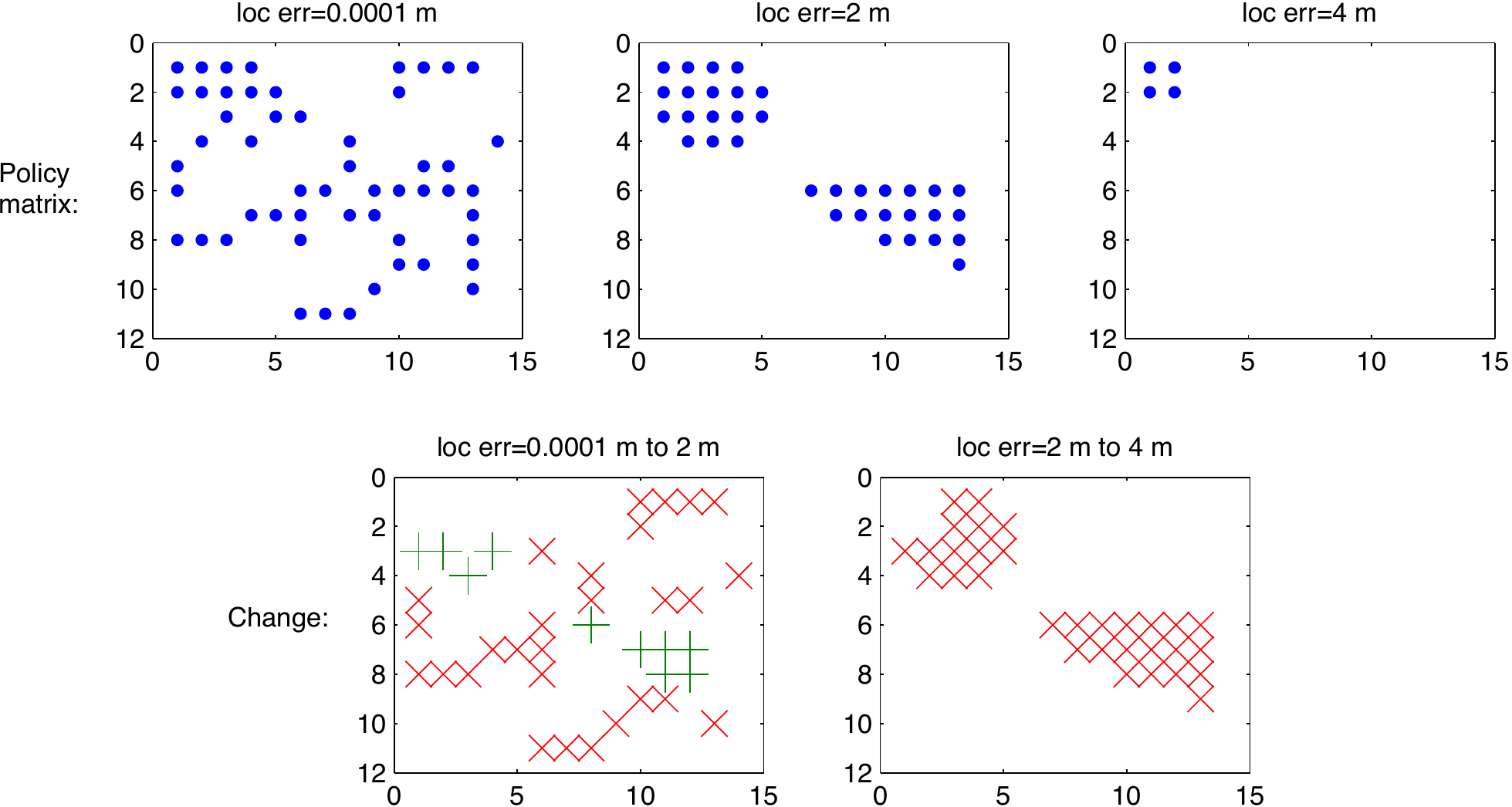}
    \caption{Optimised relaying policy, and the change in optimised relaying policy for selected location accuracies. Blue dots in the policy matrix show the geographic points where relaying is used -- direct transmissions are used in the rest. In the \textit{change} matrix, green pluses are added relaying points and red crosses are removed relaying points.}
\label{fig:mmPr_loc_err_policies_changes}
\end{figure*}
%

Since the ray-tracing based data set introduces a high level of variability in the policy decision process, we have looked beyond the ''fraction of points where relaying is preferred'' to study specifically what happens to the relay policies. An example is detailed in the plots in Fig. \ref{fig:mmPr_loc_err_policies_changes}, which shows the full relay policies for three selected location error values, as well as the change in policy between these values.

These results show that even though the fraction of points where relaying is preferred does not change much, the policy optimisation can actually imply both additions and removals of points to the relay policy. Consequently, the policy optimisation does not just generally reduce or increase the number of relaying points, but adds and removes points based on the individual conditions.


\section{Conclusions and Outlook} \label{sec:conclusion}
Location information can be useful for optimising networking functions such as relaying. As location data is needed for each mobile node, such information can be collected by access points in linear effort with respect to number of mobile nodes, while the number of links grows quadratically. However, the localisation error and the chosen update rate of location information in conjunction with the mobility model affects the accuracy of location information and hence all these parameters need to be taken into account in the design of optimal policies.
 
This paper develops a Markov model that can capture the joint impact of localisation errors and inaccuracies of location information due to forwarding delays and mobility. The Markov model can be the tool for parametric studies and hence deployment optimisations of relaying solutions. Furthermore, the Markov model is used to develop algorithms that determine optimal location-based relay policies that take the aforementioned factors into account. Applying the model to analyse the impact of deployment parameter choices on the performance of location-based relaying in WLAN scenarios with free-space propagation conditions shows that both the increase of location errors and the increasing speed of node mobility can be compensated by higher update rates in order to maintain relaying performance. However, location errors can only be compensated by fresher location information up to some maximum, at which location information becomes useless for relay choice optimisation. Similar results are shown in an measurement-based indoor office scenario with more complex mobility model and multipath propagation.

The presented analysis framework is particularly useful for off-line deployment optimisation, where the determined optimal relay policies can be stored in the access point and an appropriate location information update rate can be determined for the used localisation system. Alternatively, the benefit of using a more accurate localisation system can be analysed and based on the results, one can decide if such an upgrade is worthwhile.

While the examples and the model setup in the paper optimises relaying policies with respect to throughput, other target metrics can be easily included as long as those are determined by the geographic positions of the two end-points of the transmission link. For instance, the target metric could also take distance-depen\-dent energy budgets of the transmission by the relay nodes into account, i.e. use a weighted combination of throughput maximization and relay transmission energy minimization. Such choice of optimisation metric can increase network lifetime. For explicit optimisation of network lifetime, further extensions of the Markov model can be investigated to also include uplink forwarding of energy levels of nodes to the access point.

\begin{acknowledgements}
This work has been performed in the framework of the ICT project ICT-248894 WHERE2, which is partly funded by the European Union. The Telecommunications Research Center Vienna (FTW) is supported by the Austrian Government and by the City of Vienna within the competence center program COMET.
\end{acknowledgements}


\bibliographystyle{spmpsci}      

\end{document}